\documentclass{article}

\usepackage{xcolor}
\definecolor{comments}{rgb}{0.0, 0.0, 0.0}
\definecolor{recent}{rgb}{0.0,0.0,0.0}
\definecolor{mc}{rgb}{0,0,0} 
\definecolor{mine}{rgb}{0,0,0}
\usepackage{amsmath} 
\usepackage{amssymb}
\usepackage{textcomp}
\usepackage{bm}
\usepackage{graphicx}
\usepackage{graphics}
\usepackage{subcaption}
\usepackage[symbol]{footmisc}

\usepackage[sort&compress, numbers]{natbib}
\allowdisplaybreaks
\usepackage{PRIMEarxiv}

\usepackage[utf8]{inputenc} 
\usepackage[T1]{fontenc}    
\usepackage{hyperref}       
\usepackage{url}            
\usepackage{booktabs}       
\usepackage{amsfonts}       
\usepackage{nicefrac}       
\usepackage{microtype}      
\usepackage{lipsum}
\usepackage{fancyhdr}       
\usepackage{graphicx}       
\graphicspath{{media/}}     

\title{Sparse Bayesian neural networks for regression: Tackling overfitting and computational challenges in uncertainty quantification
}

\author{
  Nastaran Dabiran \\
  Department of Civil and Environmental Engineering \\
  Carleton University  \\
  Ottawa, ON, Canada \\
   \\
   \And
  Brandon Robinson \\
  Department of Civil and Environmental Engineering \\
  Carleton University  \\
  Ottawa, ON, Canada \\
   \\
   \And
   Rimple Sandhu \\
   Computational Science Center \\
   National Renewable Energy Laboratory \\
    Golden, CO, United States\\
   \And
   Mohammad Khalil\thanks{\textit{Sandia National Laboratories is a multimission laboratory managed and operated by National Technology \& Engineering Solutions of Sandia, LLC, a wholly owned subsidiary of Honeywell International Inc., for the U.S. Department of Energy’s National Nuclear Security Administration under contract DE-NA0003525. This paper describes objective technical results and analysis. Any subjective views or opinions that might be expressed in the paper do not necessarily represent the views of the
U.S. Department of Energy or the United States Government.}} \\
   Quantitative Modeling \& Analysis Department \\
   Sandia National Laboratories\\
   Livermore, CA, United States \\
   \And
   Dominique Poirel \\
   Department of Mechanical and Aerospace Engineering\\
    Royal Military College of Canada\\  
    Kingston, ON, Canada\\
    \\
    \And
    Abhijit Sarkar\\
    Department of Civil and Environmental Engineering\\
    Carleton University\\
    Ottawa, ON, Canada\\
}

\begin{document}
\maketitle
\begin{abstract}
Neural networks (NNs) are primarily developed within the frequentist statistical framework. Nevertheless, frequentist NNs lack the capability to provide uncertainties in the predictions, and hence their robustness can not be adequately assessed. Conversely, the Bayesian neural networks (BNNs) naturally offer predictive uncertainty by applying Bayes' theorem. However, their computational requirements pose significant challenges. Moreover, both frequentist NNs and BNNs suffer from overfitting issues when dealing with noisy and sparse data, which render their predictions unwieldy away from the available data space. To address both these problems simultaneously, we leverage insights from a hierarchical setting in which the parameter priors are conditional on hyperparameters to construct a BNN by applying a semi-analytical framework known as nonlinear sparse Bayesian learning (NSBL). We call our network \textit{sparse Bayesian neural network} (SBNN) which aims to address the practical and computational issues associated with BNNs.  Simultaneously, imposing a sparsity-inducing prior encourages the automatic pruning of redundant parameters based on the automatic relevance determination (ARD) concept.  This process involves removing redundant parameters by optimally selecting the precision of the parameters prior probability density functions (pdfs), resulting in a tractable treatment for overfitting. To demonstrate the benefits of the SBNN algorithm, the study presents an illustrative regression problem and compares the results of a BNN using standard Bayesian inference, hierarchical Bayesian inference, and a BNN equipped with the proposed algorithm. \textcolor{mine}{Subsequently, we demonstrate the importance of considering the full parameter posterior by comparing the results with those obtained using the Laplace approximation with and without NSBL.} 
\end{abstract}

\keywords{Sparse Bayesian learning, Nonlinear sparse Bayesian learning, Hierarchical Bayesian inference, Bayesian neural network.}

\section{Introduction}
\label{sec:intro} 
In the realm of neural networks, the frequentist statistical framework \textcolor{mine}{\cite{konishi2008information, murphy2012}, which views probability as the frequency of an event in the limit of infinite trials,} has been the dominant paradigm. However, despite their remarkable accomplishments, frequentist NNs have certain limitations. One significant limitation is their inability to provide meaningful estimates of uncertainty in their predictions. This limitation has stimulated a notable shift in recent years towards exploring Bayesian statistical formulations of neural networks \cite{mackay1992, neal1992, mackay1995, neal1996}. This makes NNs capable of providing both predictions and the uncertainty associated with those predictions. While incorporating prediction uncertainty is advantageous, the computational requirements pose significant challenges in BNNs. \textcolor{mine}{ In deep BNNs, the Bayesian posterior is high-dimensional and highly} \textcolor{mc}{non-Gaussian}. Hence, the \textcolor{comments}{parameter posteriors and} predictive distribution integral \textcolor{comments}{are} not \textcolor{comments}{available} in the closed form \cite{izmailov2021}. To tackle this challenge, two \textcolor{mc}{main strategies are}: (1) Markov-Chain Monte-Carlo (MCMC) and (2) Variational Inference \cite{jospin2022hands}. 

Despite their limitations, MCMC techniques are widely considered as one of the most effective and prevalent solutions for obtaining samples from the posterior distributions in Bayesian statistics. The most applicable MCMC methods for BNNs are the Metropolis-Hastings (MH) \cite{chib1995, wilson2020} and Hamiltonian Monte Carlo algorithm (HMC) \cite{neal2011, jospin2022hands}. \textcolor{mine}{While MCMC algorithms are capable of sampling from the posterior pdfs, their lack of scalability and computational challenges render them less popular for deep BNNs, given the large number of parameters in the network.}
Conversely, variational Bayesian (VB) inference \cite{blei2017,graves2011practical, shridhar2019comprehensive} provides a faster and more computationally efficient approach than traditional MCMC and has gained popularity in recent years as an approximation method for BNNs. VB aims to find a parametrized approximation of the posterior, referred to as the variational distribution, that is as close as possible to the exact posterior by optimizing its parameters \cite{shridhar2019comprehensive}. Some practical implementations of VB for NNs known as Bayes by Backprop \cite{blundell2015}, Bayesian Hypernetwork \cite{kruegeret2018}, Bayesian Layers \cite{tran2019}. VB approaches typically provide unimodal Gaussian approximation to multimodal posterior \cite{izmailov2021}, which may result in inaccurate approximations of complex probability distributions, particularly in scenarios where the true posterior is not unimodal. 

Besides the aforementioned issues, a large number of parameters \textcolor{mc}{(e.g., weights and biases)} in NNs render them prone to overfitting, which adversely affects their generalization capabilities \cite{szegedy2013}. Many attempts have been made to alleviate this issue by applying pruning techniques \textcolor{mine}{\cite{vadera2022methods}} to reduce the size of the network by removing redundant parameters such as \textcolor{mc}{optimal brain surgeon (OBS), which applies network-wise pruning threshold on a trained NN \cite{hassibi1993}. Moreover,} dropout  \cite{srivastava2014dropout} \textcolor{mc}{and dropconnect \cite{wan2013regularization} are also effective regularization techniques to reduce overfitting,  by either dropping states of neural units (i.e., set to zero) or dropping weights of synaptic connections
 in  the training process respectively,} forcing the network to learn more robust representations \cite{sakai2019}. Despite their convenience in implementation, they may lack some expressivity and less effective in capturing the predictive uncertainty \cite{chan2020unlabelled, jospin2022hands}. 

Previously, MacKay \cite{mackay1995} suggested finding a sparse solution by applying the idea of ARD priors, known as evidence framework, where the integral of evidence (or marginal likelihood) is approximated using Laplace approximation around the MAP estimation of parameters. However, Laplace approximation \textcolor{mine}{involves solving a second-order optimization problem, which can be computationally intensive for high-dimensional models. While efforts have been made to enhance the efficiency of Hessian matrix calculations \cite{zhou2022sparse}, it is important to note the inherent Gaussian assumption of the posterior in the vicinity of MAP estimate built in  Laplace approximation. In cases where the true posterior is significantly non-Gaussian, this approximation can introduce significant errors, thereby arriving at suboptimal representation.}

To the best of our knowledge, none of the available methods have been assessed on their capability to match the posterior parameters distribution and predictive distribution in the hierarchical setting, even though they have been demonstrated to deliver improved predictions or more accurate uncertainty estimates. Moreover, many of them are restricted to the choice of prior and are not equipped with sparsity-inducing apparatus. We leverage recent insights from approximation techniques in a hierarchical setting to construct a BNN by applying a semi-analytical framework known as nonlinear sparse Bayesian learning (NSBL) \cite{sandhu2020model,sandhu2021}. NSBL employs the use of a so-called hybrid prior, allowing a combination of ARD priors and informative priors to be used, as well as leveraging a Gaussian mixture model (GMM) approximation of  the likelihood function times the known prior, providing semi-analytical  (as functions of the hyperparameters) expressions of the model evidence, the parameter posterior, the objective function, its gradient vector and Hessian matrix. This setup provides a good choice for models (e.g., BNNs) consisting of non-Gaussian prior as well as non-Gaussian likelihood \cite{dabiran_jcp2023}. \textcolor{mc}{In this paper, we  apply} NSBL on BNNs, which addresses both overfitting and computational challenges in computing parameter posterior. We refer to our algorithm as \textit{sparse Bayesian neural network} (SBNN). This algorithm will allow for an optimal sparse NN to be identified, which is nested within a fully connected network by removing redundant weight parameters, providing a tractable treatment for overfitting through an evidence optimization algorithm. 

The use of sparsity-inducing priors in NNs is not novel in itself \cite{mackay1995,zhou2022sparse}. However, the NSBL methodology provides an efficient mechanism by which a GMM approximation of the parameter posterior pdf is obtained as a function of the hyperparameters \cite{sandhu2021,sandhu2020model}. This permits the semi-analytical calculation of many quantities of interest within the SBNN framework, thereby alleviating computational complexities encountered in BNNs. Optimizing model evidence with respect to the hyperparameters, the SBNN algorithm identifies the optimal sparse NN structure for the given dataset. \textcolor{mine}{Currently, the methodology for SBNNs relies on sampling. Thus, when employed for deep NNs, it can become prohibitively expensive due to the high dimensional parameter space. Opting for alternative approaches, such as \textit{last layer learning} \cite{schwobel2022last} powered by NSBL, could present a viable solution. In this approach, the computation of Bayesian quantities is required exclusively in the final layer \cite{wilson2016deep, van2018learning}. However, this approach of only training the last layer in a Bayesian setting may not be appropriate in all situations}. \textcolor{mine}{Alternatively, \textit{layer-wise learning} \cite{hinton2006fast,bengio2006greedy}, whereby the network trains one layer at a time also provides a good environment for deploying the SBNN methodology for deep NNs. Furthermore, applying a sampling-free approach \cite{bridgman2023robust} for the GMM construction of the likelihood function enhances computational efficiency in the current algorithm.}

 For the sake of completeness, first, we review the mathematical definition of frequentist and Bayesian neural networks. Next, \textcolor{comments}{the concept of sparse representation of BNN is introduced} together with the required changes to the Bayesian setting. This is followed by a review description of the NSBL algorithm. Finally, we apply NSBL to a regression example to infer the weight and bias parameters of a shallow neural network. These numerical investigations demonstrate the application of this algorithm to identify the optimal sparse neural network, which reduces overfitting. Simultaneously, it highlights the accuracy of NSBL in comparison with \textcolor{mine}{standard Bayesian inference, }hierarchical Bayesian inference, and with the Laplace approximation with and without NSBL.

\section{Frequentist and Bayesian Approaches \\ for Neural Networks}
\noindent  
The structure of a NN fundamentally consists of processing units, commonly referred to as neurons allocated in the layers, which are connected through weight and bias parameters \cite{higham2019deep}. 
At a high level, the difference between a standard NN and a BNN is that the trained network will have optimized point estimates for the parameters in a frequentist framework, whereas in a BNN, posterior pdfs are obtained through Bayesian approaches \cite{mackay1992,mackay1995}. 

\subsection{Neural Network Notation}
\label{sec.2.1}
At its core, a NN is simply an arbitrary functional mapping of an input vector $\mathbf{x}  \in \mathbb{R}^{N_\mathrm{x}}$ to an output entity $\mathbf{y} \in \mathbb{R}^{N_\mathrm{y}}$. Traditionally, NNs are built using one input layer $l_0$ and several successive hidden layers $l_i, i= 1,2,...,n-1$ of a linear transformation (given by unknown parameter vector $\mathbf{\phi} \in \mathbb{R}^{N_\phi}$) equipped with element-wise non-linear activation functions, and finally the output layer $l_n$. Here, $\phi= \{W_{ji}^{[l]},b_i^{[l]}\}$, where $W_{ji}^{[l]}$  is the weight of the network connections from neuron $i$ to neuron $j$ associated to the layer $l$ and $b_i^{[l]}$ defines the bias parameter of neuron $i$ in the layer $l$ \cite{higham2019}. The simplest architecture of neural networks can be formalized as
\begin{equation} \label{eq:nn}
\begin{aligned}
l_0 &= \mathrm{x}\\
l_i &= \sigma (W_{ji}^{[l]} l_{i-1}+ b_i^{[l]})\\
y &= l_n\\
\end{aligned}
\end{equation}
where $\sigma$ represents the activation function which nonlinearly maps the input to the output. A few commonly well-known activation functions are \textcolor{comments}{sigmoid, hyperbolic tangent (tanh), and rectified linear unit (Relu)} \cite{higham2019}.  

\subsection{Making a neural network Bayesian}
\label{sec.2.3}
In contrast to the frequentist neural network, which we find a point estimate of the parameters $\phi$, in the BNNs we obtain posterior distributions of parameters. Following the standard Bayesian formulation, the expression for the parameter posterior pdf is written as

\begin{equation} \label{2:posterior_standard}
\mathrm{p}(\bm{\phi} \vert \mathcal{D}) = \frac{\mathrm{p}(\mathcal{D}\vert\bm{\phi})\mathrm{p}(\bm{\phi})}{\mathrm{p}(\mathcal{D} ) }
=  \frac{\mathrm{p}(\mathcal{D}\vert\bm{\phi})\mathrm{p}(\bm{\phi})}{\int \mathrm{p}(\mathcal{D}\vert\bm{\phi})\mathrm{p}(\bm{\phi}) d\bm{\phi} }
\end{equation}

where $\mathrm{p}(\mathcal{D}\vert\bm{\phi})$ is the likelihood function assumed to be known for a given value of $\boldsymbol\phi$, $\mathrm{p}(\mathcal{D})$ is the model evidence, and $\mathrm{p}(\boldsymbol{\phi})$ is parameter prior pdf, which \textcolor{comments}{encodes the} known information. A typical selection of prior distribution in Bayesian analysis involves Gaussian distribution, denoted as $\mathrm{p}\left(\boldsymbol{\phi}\right) = \mathcal{N}(0,\mathbf{A}^{-1})$, \textcolor{comments}{where $\mathbf{A}$ is the precision matrix (inverse of covariance matrix) and the mean vector being zero}; and Gaussian prior and Gaussian likelihood satisfy conjugact property  \cite{fink1997compendium}. \textcolor{comments}{The precision} matrix $\mathbf{A}=\operatorname{Diag}(\boldsymbol{\alpha})$, where $\boldsymbol{\alpha}$ is assumed to be a known constant, \textcolor{comments}{representing independence among the elements  of $\boldsymbol{\phi}$}. The posterior distribution captures our uncertainty about the true value of the parameters given some (presumably noisy, sparse, and incomplete) training data $\mathcal{D}$.

\subsection{Automatic sparsity in Bayesian neural networks}
\label{sec2.2.1}
Generally, increasing the number of model parameters can improve the data-fitting capability of a given model. However, an over-parameterized model suffers from \textit{overfitting}. An overfitted model lacks the generalization capability to unforeseen data and \textcolor{comments}{leads to large inconsistencies in the predictions} \cite{guo2017calibration, nixon2019measuring}. One way to tackle this problem is to introduce a higher level of the hierarchy in the Bayesian setting whereby the prior is conditional on the hyperparameter and incorporates sparsity-inducing priors \cite{murphy2012, tipping2001}.

\subsubsection{ Hierarchical Bayesian inference}\label{sec:2.3.1}
Assuming a one-level hierarchy and applying Bayes' theorem, the joint posterior of the parameters $\boldsymbol{\phi}$ and hyperparameters $\boldsymbol{\alpha}$ is written as 

\begin{equation} \label{2:posterior}
\mathrm{p}(\bm{\phi},\bm{\alpha} \vert \mathcal{D}) = \frac{\mathrm{p}(\mathcal{D}\vert\bm{\phi})\mathrm{p}(\bm{\phi} \vert \bm{\alpha})\mathrm{p}(\bm{\alpha})}{\mathrm{p}(\mathcal{D}) },
\end{equation}
where the likelihood function $\mathrm{p}(\mathcal{D}\vert\bm{\phi})$ is assumed to be known for \textcolor{comments}{a given} value of $\boldsymbol\phi$. Note that the likelihood function is indirectly influenced by $\bm{\alpha}$ through the conditional dependence of the parameters $\bm{\phi}$ on the hyperparameters $\bm{\alpha}$. A key requirement for computing the posterior in this hierarchical Bayesian setup is the specification of the prior $\mathrm{p}(\boldsymbol{\phi} \mid \boldsymbol{\alpha})$ as will be discussed later. Finally, $\mathrm{p}(\mathcal{D})$ in the denominator denotes model evidence or marginal likelihood. When the model evidence becomes too complex to calculate directly (i.e., involves high-dimensional integral), one typically resorts to using sampling methods \cite{sandhu2021,chib1995}.  MCMC sampling methods  construct a Markov chain, a sequence of random samples, \textcolor{comments}{whose stationary distribution mimics parameter posterior} \cite{gilks1995markov}. A well-known extension to classical MCMC \textcolor{comments}{for sampling complex} multimodal posterior pdfs and \textcolor{comments}{estimating the associated} model evidence, is known as transitional MCMC (TMCMC) \cite{ching2007}. Although sampling approaches are reliable since they directly sample from the unnormalized posterior, they are computationally expensive and slow to converge due to the large number of parameters in deep BNNs.

\subsubsection{ Nonlinear sparse Bayesian learning}
\label{sec:2.3.2}

The computational challenges associated with the sampling in hierarchical Bayesian setup inspired the development of sparse Bayesian learning (SBL) (also known as relevance vector machine (RVM)) \cite{tipping2001,drugowitsch2013}, whereby Eq. \ref{2:posterior_standard} can be rewritten as

 \begin{equation} \label{2:inference}
\begin{aligned}
\mathrm{p}(\boldsymbol{\phi} \mid \mathcal{D}, \boldsymbol{\alpha})=\frac{\mathrm{p}(\mathcal{D} \mid \boldsymbol{\phi}) \mathrm{p}(\boldsymbol{\phi} \mid \boldsymbol{\alpha})}{\mathrm{p}(\mathcal{D} \mid \boldsymbol{\alpha})}=\frac{\mathrm{p}(\mathcal{D} \mid \boldsymbol{\phi}) \mathrm{p}(\boldsymbol{\phi} \mid \boldsymbol{\alpha})}{\int \mathrm{p}(\mathcal{D} \mid \boldsymbol{\phi}) \mathrm{p}(\boldsymbol{\phi} \mid \boldsymbol{\alpha}) d \boldsymbol{\phi}},
\end{aligned}
\end{equation}

where we assume $\boldsymbol{\phi}$ is assigned a Gaussian Automatic Relevance Determination (ARD) prior of the form $\mathrm{p}\left(\boldsymbol{\phi} \mid \boldsymbol{\alpha}\right)=\mathcal{N}\left(\boldsymbol{\phi} \mid \mathbf{0}, \mathbf{A}^{-1}\right)$ where the precision matrix $\mathbf{A}=\operatorname{Diag}(\boldsymbol{\alpha})$. This choice of ARD prior encourages the automatic pruning of redundant parameters \cite{murphy2012, tipping2001}; The hyperparameter $\alpha_i$ dictates the complexity of the model by controlling the contribution of parameter $\phi_i$ \cite{sandhu2021,sandhu2020model}. In this setting, $\mathrm{p}(\boldsymbol{\alpha})$ is the hyperprior pdf and is typically chosen as Gamma distribution \textcolor{comments}{(with independent components)} of the form $\mathrm{p}(\boldsymbol{\alpha})=\prod_{i=1}^{N_\alpha} \operatorname{Gamma}\left(\alpha_i \mid s_i, r_i\right) = \prod_{i=1}^{N_\alpha} \frac{r_i^{s_i}}{\Gamma\left(s_i\right)} \alpha_i^{s_i} e^{-r_i \alpha_i}$ with the known shape parameter $s_i$ and rate parameter $r_i$ \cite{tipping2001}.
The limit case of $s_i \rightarrow 0$ and $r_i \rightarrow 0$ is of particular interest in this paper. Setting these parameters to zero leads to a Jeffreys prior $p(\log\alpha_i) \propto 1$ for the hyperparameters. Jeffreys prior \cite{jeffreys1946} is a noninformative prior and exhibits flatness over $\log \alpha_i$. This choice of hyperprior enforces the requirement of positivity in the precision parameters $\boldsymbol{\alpha}$, reducing the objective function to the log-evidence, leading to the type-II maximum likelihood estimation \cite{tipping1999, tipping2001, murphy2012}. 

While SBL has been utilized to introduce sparsity in linear-in-parameter models such as regression problems \cite{murphy2012}, it is not applicable to the case of nonlinear-in-parameter models where observations have nonlinear relationships with unknown parameters such as NNs. In such instances, the semi-analytical framework known as nonlinear sparse Bayesian learning (NSBL) \textcolor{comments}{has been recently introduced to address the computational challenges associated with hierarchical Bayesian inference} addresses these practical issues by employing an approximate to the parameter posterior pdfs. NSBL setup permits the inclusion of hybrid priors of the form  $\boldsymbol{\phi}= \{\boldsymbol{\phi}_\alpha, \boldsymbol{\phi}_{-\alpha}\}$ \cite{sandhu2017,sandhu2020model,sandhu2021,sandhu2022}. While $\boldsymbol{\phi}_{-\alpha} \in \mathbb{R}^{N_\phi-N_\alpha}$ contains parameters that are \textit{a priori} relevant and have a prescribed prior, $\boldsymbol{\phi}_\alpha \in \mathbb{R}^{N_\alpha}$ is defined as the set of parameters whose relevance is \textit{a priori} unknown and assigned ARD prior. Therefore, the joint prior pdf of $\boldsymbol{\phi}$ is denoted as
$\mathrm{p}(\boldsymbol{\phi} \mid \boldsymbol{\alpha}) = \mathrm{p}(\boldsymbol{\phi}_{-\alpha}) \mathrm{p} ({\boldsymbol\phi}_{\alpha}\mid \boldsymbol{\alpha}) = \mathrm{p}(\boldsymbol{\phi}_{-\alpha}) \mathcal{N}\left(\boldsymbol{\phi} \mid \mathbf{0}, \mathbf{A}^{-1}\right)$. Given the hybrid prior, Eq. \ref{2:inference} can be written as,

\begin{equation} \label{2:hybrid-posterior}
\begin{aligned}
\mathrm{p}(\boldsymbol{\phi} \mid \mathcal{D}, \boldsymbol{\alpha})=\frac{\mathrm{p}(\mathcal{D} \mid \boldsymbol{\phi}) \mathrm{p}(\boldsymbol{\phi}_{-\alpha}) \mathcal{N}\left(\boldsymbol{\phi_{\alpha}} \mid \mathbf{0}, \mathbf{A}^{-1}\right)}{\mathrm{p}(\mathcal{D} \mid \boldsymbol{\alpha})}\\
\propto {\mathrm{p}(\mathcal{D} \mid \phi) \mathrm{p}\left(\phi_{-\alpha}\right)} \mathcal{N}\left(\phi_{\alpha} \mid \mathbf{0}, \mathbf{A}^{-1}\right).
\end{aligned}
\end{equation}

Assigning ARD priors $\mathrm{p}(\mathbf{\phi}_\alpha \mid \bm{\alpha})$ permits the automatic pruning of redundant parameters \textcolor{comments}{(explained later)}, while $\mathrm{p}(\mathbf{\phi}_{-\alpha})$ permits the inclusion of prior information about certain model parameters. NSBL constructs a GMM approximation of the likelihood times prior pdf of \textit{a priori} relevant parameters ${\mathrm{p}(\mathcal{D} \mid \phi) \mathrm{p}\left(\phi_{-\alpha}\right)}$ \cite{sandhu2021,sandhu2022}

\begin{equation} \label{2:GMM}
\mathrm{p}(\mathcal{D} \mid \phi) \mathrm{p}\left(\phi_{-\alpha}\right) \approx \sum_{k=1}^{K} a^{(k)} \mathcal{N}\left(\phi \mid \boldsymbol{\mu}^{(k)}, \boldsymbol{\Sigma}^{(k)}\right),
\end{equation}

where $K$ denotes the total number of kernels, $a^{(k)} \in \mathbb{R}$ is the kernel coefficient $(a^{(k)}>0, \sum_k^K a^{(k)} = 1)$ and $\left.\mathcal{N}(\boldsymbol{\phi} \mid \boldsymbol{\mu}^{(k)}, \boldsymbol{\Sigma}^{(k)})\right.$ is a Gaussian pdf with mean vector $\boldsymbol{\mu}^{(k)} \in \mathbb{R}^{N_\phi}$ and covariance matrix $\boldsymbol{\Sigma}^{(k)} \in \mathbb{R}^{N_\phi \times N_\phi}$ \cite{sandhu2021}. 
This GMM is an entity independent of $\boldsymbol{\alpha}$ parameter; hence, there is no need to recompute it as the algorithm iterates through different values of $\boldsymbol{\alpha}$ during optimization \textcolor{comments}{to find the MAP estimation of $\boldsymbol{\alpha}$}.  \textcolor{comments}{As the ARD prior $\mathrm{p}(\mathbf{\phi}_\alpha \mid \bm{\alpha})$ is Gaussian, the approximation in Eq. \ref{2:GMM} leads to a Gaussian mixture representation of the parameter posterior given by } 

\textcolor{mine}{\begin{equation} \label{2:GMM-paramposterior}
\hat{\mathrm{p}}(\mathbf{\phi}\mid \mathcal{D},\boldsymbol{\alpha}) = \frac{\sum_{k=1}^{K} a^{(k)} \mathcal{N}\left(\phi \mid \boldsymbol{\mu}^{(k)}, \boldsymbol{\Sigma}^{(k)}\right)\mathcal{N}\left(\phi_{\alpha} \mid \mathbf{0}, \mathbf{A}^{-1}\right)}{\hat{\mathrm{p}}(\mathcal{D} \mid \boldsymbol{\alpha})}  ,
\end{equation}
where $\hat{\mathrm{p}}(\mathcal{D} \mid \boldsymbol{\alpha})$ is an estimate of model evidence obtained by substituting Eq. \ref{2:GMM} and ARD prior $\mathcal{N}\left(\phi_{\alpha} \mid \mathbf{0}, \mathbf{A}^{-1}\right)$ in $
\mathrm{p}(\mathcal{D} \mid \boldsymbol{\alpha})=\int \mathrm{p}(\mathcal{D} \mid \boldsymbol{\phi}) \mathrm{p}(\boldsymbol{\phi} \mid \boldsymbol{\alpha}) d \boldsymbol{\phi}$.} The use of \textcolor{comments}{GMM provides} semi-analytical expressions for many entities of interest, such as the model evidence and objective function, and the gradient and Hessian of the objective function (more details on mathematical derivation can be found in \cite{sandhu2021, sandhu2022}). Moreover, the GMM relaxes the Gaussian assumptions in both the likelihood function, the known prior \textcolor{comments}{and posterior} \cite{sandhu2020model,sandhu2021}. This property of the \textcolor{comments}{NSBL algorithm offers the ability to handle (non-Gaussian)} multimodal or skewed priors and likelihood functions in BNNs. 
\textcolor{comments}{While the joint posterior of the} parameters $\bm{\phi}$ and hyperparameters $\bm{\alpha}$ are obtained within the hierarchical Bayesian framework, \textcolor{comments}{only the MAP estimation of $\boldsymbol{\alpha}$ and joint pdf of $\bm{\phi}$ are required for NSBL}. 


To induce sparsity, \textcolor{mine}{the algorithm seeks to prune redundant or irrelevant parameters} within the set of questionable parameters ${\boldsymbol{\phi}_\alpha}$. Finding the value of the hyperparameter $\boldsymbol{\alpha}$ which corresponds to the MAP estimate of the hyperparameter posterior $\mathrm{p}(\boldsymbol{\alpha} \mid \mathcal{D})$ permits the pruning of redundant parameters, \textcolor{comments}{analogous to SBL} \cite{tipping2001}. 

The optimization method of the marginal hyperparameter posterior pdf over a logarithmic scale can therefore be posed as \cite{sandhu2021,sandhu2022}

\begin{equation} \label{2:logalphaMAP}
\begin{aligned}
\log \boldsymbol{\alpha}^{\text {MAP }} 
& =\underset{\log \boldsymbol{\alpha}}{\arg \max }\{\log \mathrm{p}(\log \boldsymbol{\alpha} \mid \mathcal{D})\} ,\\
& =\underset{\log \boldsymbol{\alpha}}{\arg \max }\{\log \hat{\mathrm{p}}(\mathcal{D} \mid \log \boldsymbol{\alpha})+\sum_{i=1}^{N_\alpha}\log \mathrm{p}(\log \boldsymbol{\alpha}_i)\} ,
\end{aligned}
\end{equation}
whereby the terms independent of ${\alpha}_i$ are ignored, and the \textcolor{comments}{analytically} intractable model evidence is replaced by a GMM-based estimate of $\log \hat{\mathrm{p}}(\mathcal{D} \mid \log \boldsymbol{\alpha})$.

Once the hyperparameter MAP estimate $\log \alpha_i^{\text {MAP}}$ is determined, we use relevance indicator \cite{sandhu2021}
which is a scale-independent entity and provides a normalized metric on a scale of 0 to 1 for each of the $k$ Gaussian kernels in Eq.~\ref{2:GMM}. \textcolor{comments}{Extending the concept of relevance indicators to a multi-kernel scenario, we compute the root mean square (RMS) of $K$ relevance indicators \cite{sandhu2021}, } 
\textcolor{mine}{
\begin{equation}\label{2:relindRMS}
\begin{aligned}   
\gamma_i^{\mathrm{rms}}&=\left(\frac{1}{K} \sum_{k=1}^K\left(\gamma_i^{(k)}\right)^2\right)^{1 / 2}\\
&=\left(\frac{1}{K} \sum_{k=1}^K\left(1-\alpha_i P_{i i}^{(k)}\right)^2\right)^{1 / 2} \in[0,1],
\end{aligned}
\end{equation}}
where $P_{i i}$ is the $i$th diagonal entry of the posterior covariance matrix for kernel $k$ (see \cite{sandhu2021}) and $\alpha_i$ is the $i$th diagonal entry in the prior precision matrix $\mathbf{A}$ \cite{sandhu2022}. The parameters learned from the data will tend to have higher posterior precision compared to the prior precision \cite{sandhu2022}.

\subsection{Predictive distribution}
\label{sec2.1.3}
Once we have obtained the posterior distribution of the parameters, our focus shifts towards making predictions for a target $y^{*}$ based on an arbitrary input $x^{*}$ and the training data $\mathcal{D}$. Hence, we evaluate the predictive distribution 

\begin{equation} \label{2:prediction}
\begin{aligned}
\mathrm{p}\left(y^{*} \mid x^{*}, \mathcal{D}\right) &=\int \mathrm{p}\left(y^{*} \mid  x^{*}, \boldsymbol{\phi}\right) \mathrm{p}( \boldsymbol{\phi}, \boldsymbol{\alpha} \mid \mathcal{D})  d \boldsymbol{\phi}d \boldsymbol{\alpha}\\
\end{aligned}
\end{equation}
marginalizing over the parameters and hyperparameters in order to obtain the desired predictions.

\section{Numerical investigations}\label{sec.4}
This numerical experiment investigates how SBNN permits the data-optimal NN discovery.

\begin{figure}[ht!]
\begin{center}
\includegraphics[scale=0.5]{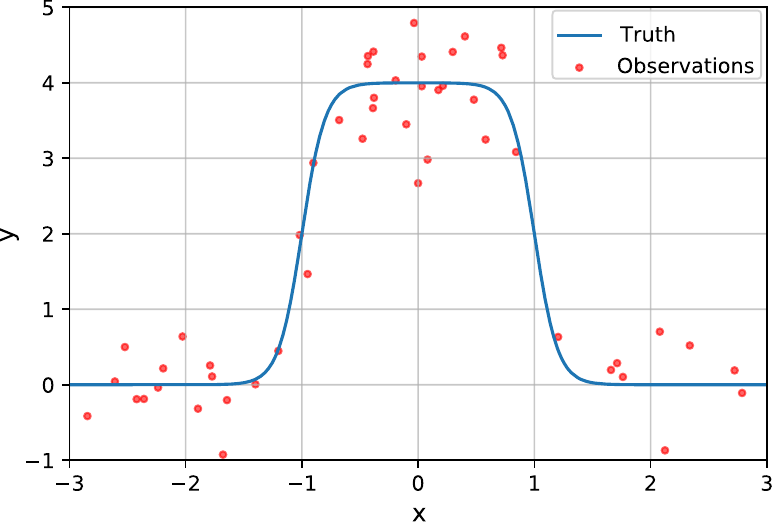}
\end{center}
\caption{Noisy observations versus the truth}
\label{data}
\end{figure}

Figure \ref{data} shows a realization of noisy measurement data generated using the boxcar function $y_i =\text{boxcar}(x_i) + \epsilon_i$ where $ \epsilon_i$ is a Gaussian random variable $\epsilon_i\sim\mathcal{N}(0,0.5)$. 

The measurements consist of 50 noisy observations equally spaced in $\boldsymbol{\mathrm{x}}\in [-3, 3]$ as shown in Figure \ref{data}. \textcolor{mine}{Boxcar} can be reasonably approximated by the sum of two hyperbolic tangent (${\tanh}$) functions as 
\begin{equation}\label{boxcar}
\text{boxcar}(\mathrm{x}) \approx \phi_1 \tanh(\phi_3 \mathrm{x} + \phi_5) + \phi_2 \tanh(\phi_4 \mathrm{x} + \phi_6 ),
\end{equation} 

\begin{figure}[!t]
\centering
\subfloat[]{\includegraphics[width=2.5in]{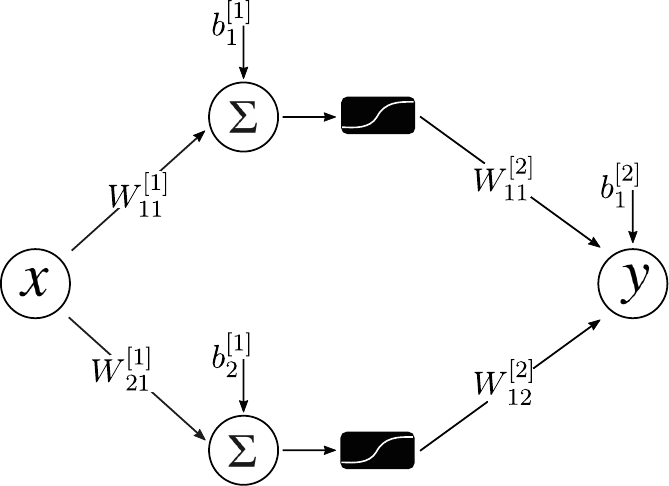}%
\label{fig:NN-2}}
\hfil\\
\subfloat[]{\includegraphics[width=2.5in]{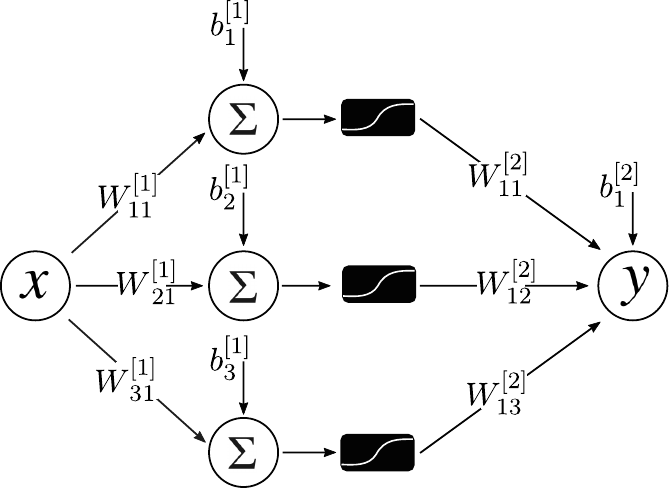}%
\label{fig:NN-3}}
\caption{Neural network architecture: (a) two-neuron NN (refer to as 1-2-1) \textcolor{comments}{(b) three-neuron NN (refer to as 1-3-1) (represents an overparameterized NN which will be used later to test the capability of NSBL in sparse representation.)}}
\label{fig:networks}
\end{figure}

\textcolor{mine}{with the} following values: $\phi_1 = \phi_2 = 2$, $\phi_3 = 5$ ,  $\phi_4 = -5$, and $\phi_5 = \phi_6 = 5$. This can be interpreted as a NN with one hidden layer and two neurons. For this NN, the activation functions in the hidden layer are considered as ${\tanh}$ shown in Figure \ref{fig:NN-2}. Recalling Eq. \ref{eq:nn}, the output of this network can be interpreted as

\begin{equation}\label{eq-2neuron}
\mathrm{y} \approx [W^{[2]}_{11} \tanh(W^{[1]}_{11} \mathrm{x} + b_1^{[1]}) + W^{[2]}_{12} \tanh(W^{[1]}_{21} \mathrm{x} + b_2^{[1]})] + b_1^{[2]}.
\end{equation}
where $W_{ij}^{[l]}$ and $b_i^{[l]}$ are the weight and bias based on the definition presented in Section \ref{sec.2.1}. Comparing Eq. (\ref{boxcar}) and \ref{eq-2neuron}, note that $b_1^{[2]}$ is zero.

\subsection{Sparse Bayesian neural network}
The process of Bayesian training of NNs involves the assignment of a prior pdf to parameters $\text{p}(\boldsymbol{\phi})$, as elaborated in Section \ref{sec.2.3}, whereby $\boldsymbol{\phi}$ includes all weight and bias parameters. If we consider non-informative prior pdfs (uniform pdf, $\text{p}(\boldsymbol{\phi})\propto 1$) for parameters $\boldsymbol{\phi}$, the posterior pdf for the parameters can be expressed as follows:

\begin{equation}\label{eq:standard_bayes_noninformative}
\text{p}(\bm{\phi} \vert \mathcal{D}) = \frac{\text{p}(\mathcal{D}\vert\bm{\phi} )\text{p}(\bm{\phi})}{\text{p}(\mathcal{D} )} \propto \text{p}(\mathcal{D}\vert\bm{\phi} )\text{p}(\phi)\propto \text{p}(\mathcal{D}\vert\bm{\phi} )
\end{equation}
\textcolor{mine}{While NSBL allows the inclusion of prior knowledge,} notice that this setting is the standard Bayesian approach where the parameters having no prior information are assigned non-informative prior.

\begin{figure*}[ht!]
\begin{center}
\includegraphics[scale=0.2]{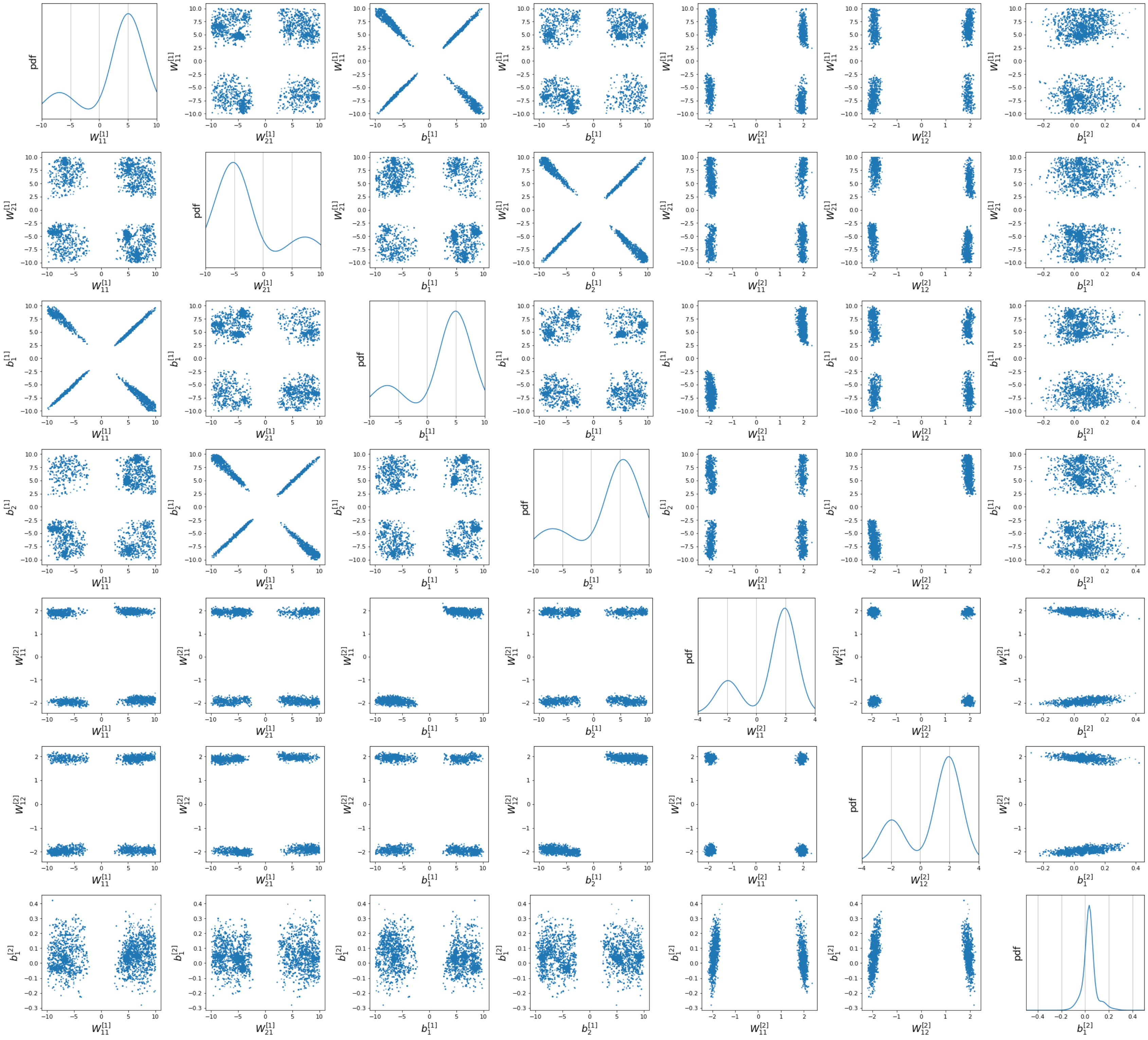}
\end{center}
\caption{Marginal and \textcolor{comments}{two-dimensional} (pairwise) joint posterior pdfs of the weight and bias parameters of the 1-2-1 NN obtained using standard Bayesian inference (This is
equivalent to the partial posterior given by the product of the known prior and the likelihood function) }
\label{fig:2neurons_standard_post}
\end{figure*}

Figure~\ref{fig:2neurons_standard_post} displays the marginal posterior pdfs and the pairwise joint TMCMC samples of the parameter posteriors \textcolor{mine}{for the 1-2-1 network}. It is worth noting that due to the symmetry in the network, Eq.~\ref{eq-2neuron} can yield identical results with various combinations of positive or negative values of parameters, a fact that is reflected by multimodal posterior pdfs. \textcolor{comments}{While} the marginal posterior distributions appear bimodal (with the exception of parameter $b_1^{[2]}$) in Figure~\ref{fig:2neurons_standard_post}, the scatterplots of the pairwise joint samples exhibit the presence of four modes \textcolor{comments}{in the pairwise joint pdfs}. \textcolor{comments}{Due to inherent symmetry in the NN in Figure~\ref{fig:NN-2}, the joint posterior of 7-dimensional parameter $\bm{\phi} = \{ W^{[1]}_{11}, b_1^{[1]},  W^{[1]}_{21}, b_2^{[1]}, W^{[2]}_{11}, W^{[2]}_{12}, b_1^{[2]} \}$ has eight modes as identified in Eq.~\ref{eq:combo1}-\ref{eq:combo8} through various combination of parameters. Note that the unequal heights of the bimodal marginal posterior pdfs are due to sampling errors.}

\begin{subequations}\label{eq:combos}
\begin{align}
        \text{Combination 1: }& \mathrm{y} = 2 \tanh(5 \mathrm{x} + 5) + 2 \tanh(-5 \mathrm{x} + 5 ) \label{eq:combo1}  \\
\nonumber (W^{[2]}_{11} = 2, W^{[2]}_{12} & = 2, W^{[1]}_{11} = 5 ,  W^{[1]}_{21} = -5,\\ 
\nonumber  b_1^{[1]} &= 5, b_2^{[1]} = 5, b_1^{[2]} = 0),\\
        \text{Combination 2: }& \mathrm{y} = 2 \tanh(-5 \mathrm{x} + 5) + 2 \tanh(5 \mathrm{x} + 5 )\label{eq:combo2} \\
\nonumber  (W^{[2]}_{11} = 2, W^{[2]}_{12} &= 2, W^{[1]}_{11} = -5 ,  W^{[1]}_{21} = 5,\\
\nonumber  b_1^{[1]} &= 5, b_2^{[1]} = 5, b_1^{[2]} = 0),  \\
        \text{Combination 3: }& \mathrm{y} = 2 \tanh(5 \mathrm{x} + 5) - 2 \tanh(5 \mathrm{x} - 5 )\label{eq:combo3} \\
\nonumber  (W^{[2]}_{11} = 2, W^{[2]}_{12} &= -2, W^{[1]}_{11} = 5 ,  W^{[1]}_{21} = 5, \\
\nonumber b_1^{[1]} &= 5, b_2^{[1]} = -5, b_1^{[2]} = 0),  \\
        \text{Combination 4: }& \mathrm{y} = 2 \tanh(-5 \mathrm{x} + 5) - 2 \tanh(-5 \mathrm{x} - 5 ) \label{eq:combo4}  \\
\nonumber  (W^{[2]}_{11} = 2, W^{[2]}_{12} &= -2, W^{[1]}_{11} = -5 ,  W^{[1]}_{21} = -5,\\
\nonumber b_1^{[1]} &= 5, b_2^{[1]} = -5, b_1^{[2]} = 0),\\
        \text{Combination 5: }& \mathrm{y} = -2 \tanh(5 \mathrm{x} - 5) + 2 \tanh(5 \mathrm{x} + 5 ) \label{eq:combo5}  \\ 
\nonumber (W^{[2]}_{11} = -2, W^{[2]}_{12} &= 2, W^{[1]}_{11} = 5 ,  W^{[1]}_{21} = 5,\\
\nonumber b_1^{[1]} &= -5, b_2^{[1]} = 5, b_1^{[2]} = 0),\\
        \text{Combination 6: }& \mathrm{y} = -2 \tanh(-5 \mathrm{x} - 5) + 2 \tanh(-5 \mathrm{x} + 5 ) \label{eq:combo6} \\
\nonumber  (W^{[2]}_{11} = -2, W^{[2]}_{12} &= 2, W^{[1]}_{11} = -5 ,  W^{[1]}_{21} = -5,\\
\nonumber b_1^{[1]} &= -5, b_2^{[1]} = 5, b_1^{[2]} = 0), \\
        \text{Combination 7: }& \mathrm{y} = -2 \tanh(5 \mathrm{x} - 5) - 2 \tanh(-5 \mathrm{x} - 5 ) \label{eq:combo7} \\
\nonumber  (W^{[2]}_{11} = -2, W^{[2]}_{12} &= -2, W^{[1]}_{11} = 5 ,  W^{[1]}_{21} = -5, \\
\nonumber b_1^{[1]} &= -5, b_2^{[1]} = -5, b_1^{[2]} = 0), \\
        \text{Combination 8: }& \mathrm{y} = -2 \tanh(-5 \mathrm{x} - 5) - 2 \tanh(5 \mathrm{x} - 5 ) \label{eq:combo8}  \\ 
\nonumber  (W^{[2]}_{11} = -2, W^{[2]}_{12} &= -2, W^{[1]}_{11} = -5 ,  W^{[1]}_{21} = 5,\\
\nonumber b_1^{[1]} &= -5, b_2^{[1]} = -5, b_1^{[2]} = 0), 
\end{align}
\end{subequations}

 In this situation, the \textcolor{mine}{1-2-1} network, illustrated in Figure \ref{fig:NN-2}, represents the optimal sparse network \textcolor{mine}{(except the bias term in the output layer $b_1^{[2]}$ which shifts the result to the left or right and it is close to zero)}.  Though not shown here for brevity, we have verified the results of NSBL applied to the optimal 1-2-1 network. If we introduce a redundant neuron, as in Figure \ref{fig:NN-3}, the number of symmetric modes in the posterior will increase. Next, we will demonstrate the automatic detection of the redundant parameters in the 1-3-1 network.

\subsubsection{\bf Nonlinear sparse Bayesian learning}

In order to assess the sparsity-inducing characteristics of the NSBL algorithm in a BNN, suppose an over-parameterized network is proposed to model the measurements by adding an additional neuron in the hidden layer whereby the overall network, as depicted in Figure~\ref{fig:NN-3}, has the following functional form, 
\begin{equation}\label{eq-overparam}
\begin{aligned}
\mathrm{y} \approx [W^{[2]}_{11} \tanh(W^{[1]}_{11} \mathrm{x} + b_1^{[1]}) + W^{[2]}_{12} \tanh(W^{[1]}_{21} \mathrm{x} + b_2^{[1]})\\
+ W_{13}^{[2]} \tanh(W^{[1]}_{31} \mathrm{x} + b_3^{[1]})]  + b_1^{[2]}.
\end{aligned}
\end{equation}

The data-generating model in Eq.~\ref{boxcar} can be recovered from Eq.~\ref{eq-overparam} by assigning zero to one set of parameters associated with either neuron one, two, or three \textcolor{mine}{ and the network will effectively become a two-neuron network.} For instance, if the algorithm recognizes the parameters associated with the third neuron \textcolor{comments}{(see Figure~\ref{fig:NN-3})} as redundant and sets them to zero, we will obtain a network with two neurons that follows Eq.~\ref{eq-2neuron}. Even if the algorithm only prunes off  $W_{13}^{[2]}$ associated with the third neuron, \textcolor{mine}{effectively eliminates the connection from the third neuron in the hidden layer to the output layer} and the same result will be obtained.

\begin{figure*}[ht!]
\begin{center}
\includegraphics[scale=0.2]{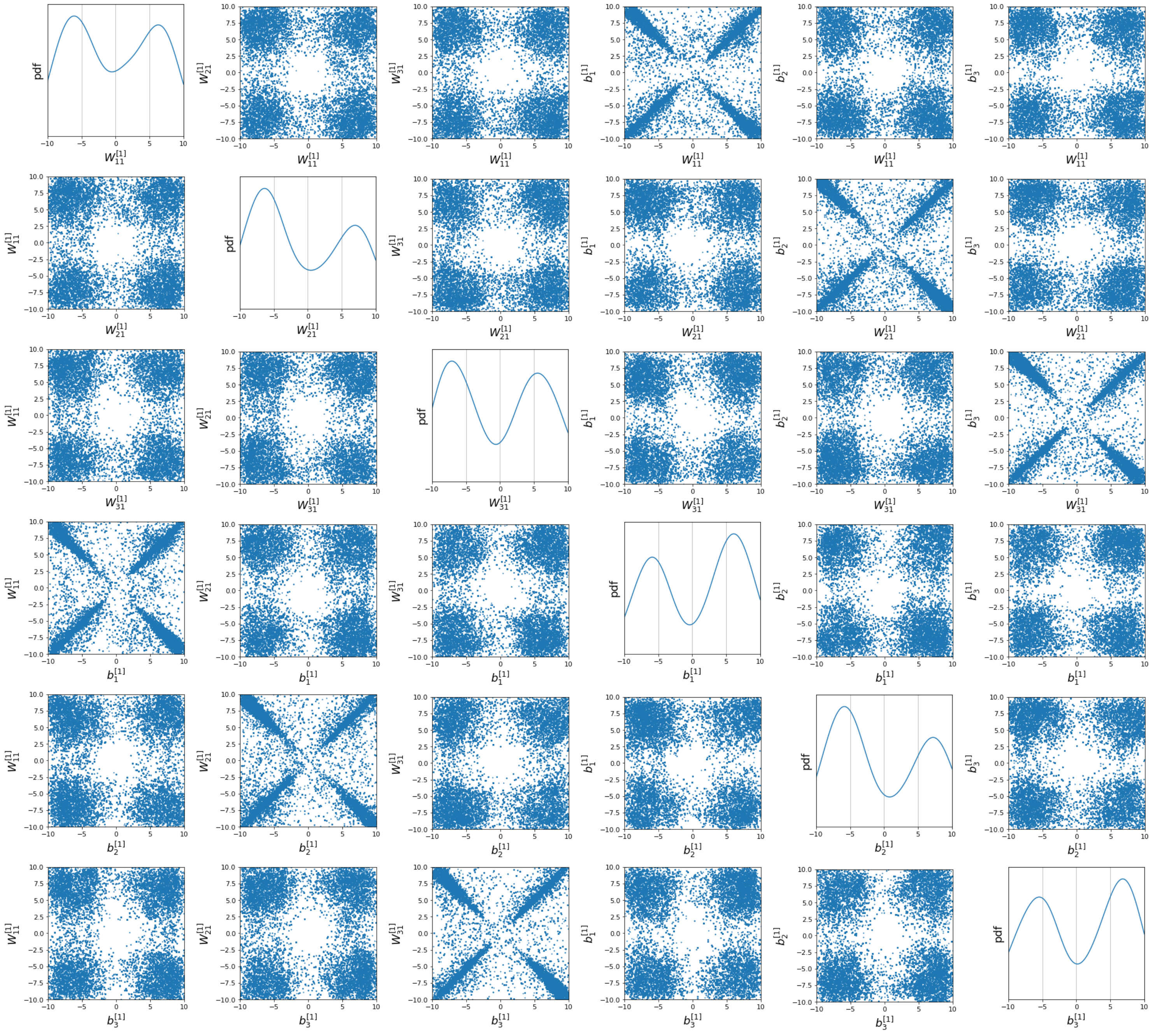}
\end{center}
\caption{Marginal posterior and \textcolor{comments}{two-dimensional} (pairwise) joint posterior of weight and bias parameters from the input layer to the hidden layer of the 1-3-1 NN obtained using standard Bayesian inference.}
\label{fig:jpdf_Bayes-lay1}
\end{figure*}

\begin{figure}[ht!]
\begin{center}
\includegraphics[scale=0.2]{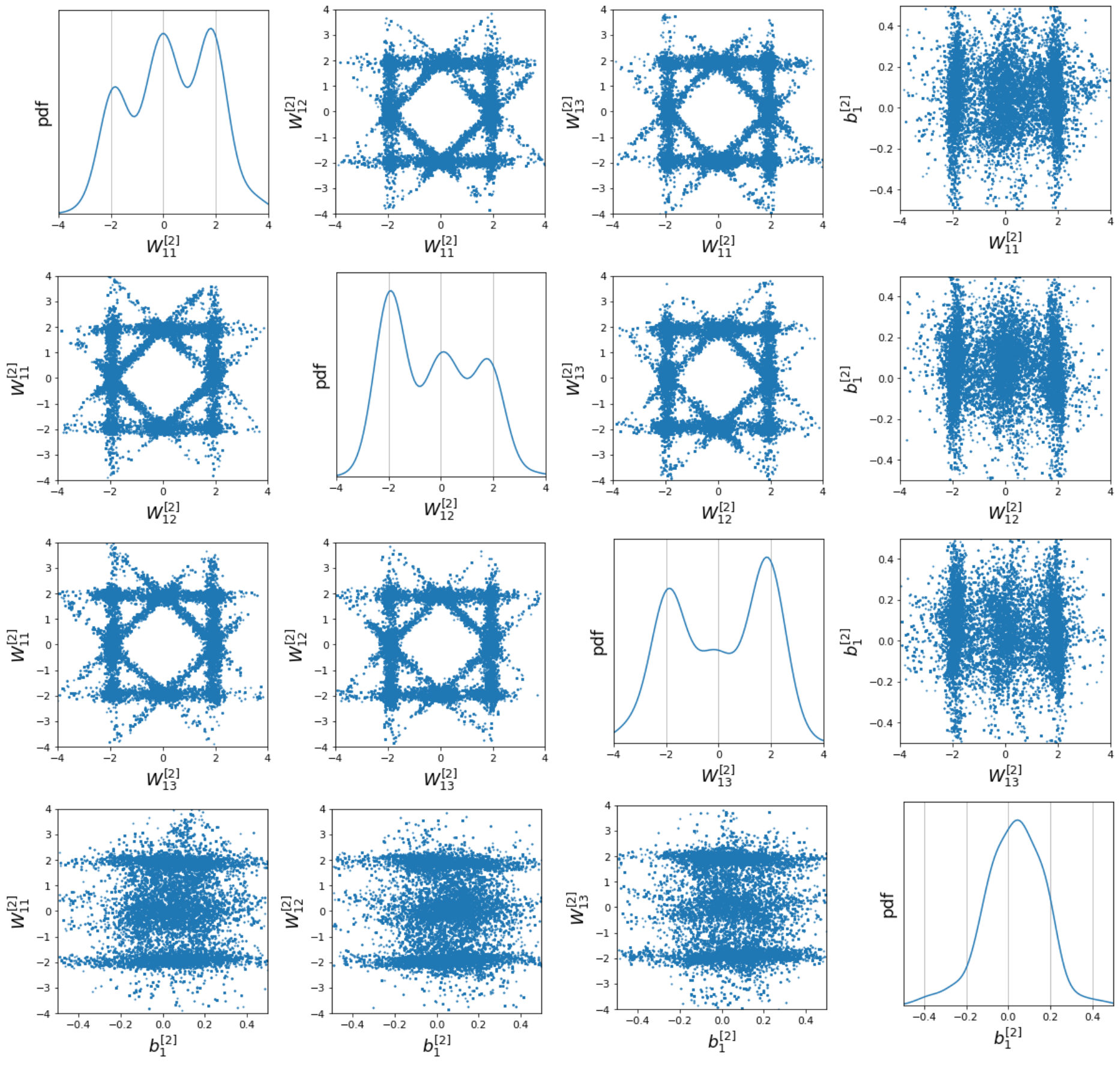}
\end{center}
\caption{Marginal posterior and joint posterior of weight and bias parameters from the hidden layer to the output layer of the 1-3-1 NN obtained using standard Bayesian inference.}
\label{fig:jpdf_Bayes-lay2}
\end{figure}

Figure~\ref{fig:jpdf_Bayes-lay1} and \ref{fig:jpdf_Bayes-lay2} depict the marginal posterior pdf and the pairwise joint TMCMC samples of the parameter posteriors for the input to the hidden layer and hidden layer to the output layer \textcolor{mine}{ of 1-3-1 NN}, respectively. Each mode in the marginal pdf plots and each cluster of samples in the joint scatter plots correspond to one combination of optimal parameters. Clearly, due to the redundant parameters in the network, more combinations \textcolor{comments}{of optimal parameters} can be constructed, and this issue is reflected in the cluster of samples in the joint scatter plots. The significant correlation between the weight and bias associated with a specific neuron is visible in \textcolor{comments}{Figures~\ref{fig:jpdf_Bayes-lay1} and \ref{fig:jpdf_Bayes-lay2}}. For instance this correlation is noticeable between $W_{i1}^{[1]}$ and $b_i^{[1]}$ for $i=1,2,3$ in Figure~\ref{fig:jpdf_Bayes-lay1}. 

In Figure~\ref{fig:jpdf_Bayes-lay2}, the joint sample plots associated with the weights connecting the hidden layer to the output suggest a strong correlation between the parameters in certain combinations manifested through the start-shaped feature. This means that changes in one parameter tend to be associated with predictable changes in another parameter and this can be proven through Eq.~\ref{eq-overparam}. \textcolor{mine}{For instance, if $W_{11}^{[2]}$ is set to $-2$, this change leads to adjustments in $b_{1}^{[1]}$ and consequently $W_{12}^{[2]}$ as  demonstrated in Eq. \ref{eq:combo5} to maintain the network's desired output.}

\begin{figure}[!t]
\centering
\subfloat[]{\includegraphics[width=2.5in]{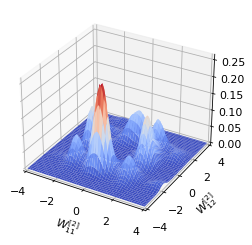}%
\label{fig:poly_objfun_sbl}}
\hfil\\
\subfloat[]{\includegraphics[width=2.5in]{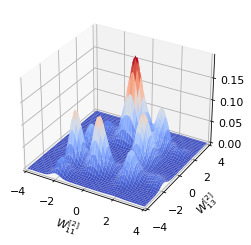}%
\label{fig:poly_logalpha_sbl}}
\hfil\\
\subfloat[]{\includegraphics[width=2.5in]{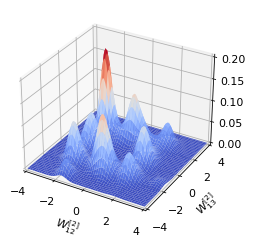}%
\label{fig:poly_relind_sbl}}
\caption{Three-dimensional surface plots of the pairwise joint parameter posterior pdfs of the weights between the hidden layer and the output layer.}
\label{fig:3d_mpost}
\end{figure}

Figure \ref{fig:3d_mpost} depicts the \textcolor{mine}{kernel density estimation (KDE) of} three-dimensional surface plots of the pairwise joint parameter posterior pdfs of the weights associated with the hidden layer to the output layers. Clearly, though samples have been generated from around each mode in the posterior, the \textcolor{comments}{number of samples} is not commensurate across all modes due to sampling error, leading to unequal heights of modes in Figure~\ref{fig:3d_mpost}.

\textcolor{comments}{The standard} Bayesian estimation results in a good data-fit \textcolor{comments}{in the regime of} training data and uncertainty during extrapolation (known as overfitting). This poor performance in extrapolation is attributed to the large uncertainty in the parameter posterior pdfs and caused by the inclusion of uniform priors \textcolor{comments}{(having large support)} along with the over-parameterized model in this setup. The overfitting issue in the over-parameterized models can be alleviated by enhancing the training data set and/or decreasing the level of noise in the data. As the current dataset is relatively small, sparse, and noisy, this fact leads to the lack of robustness in the model leading to overfiting. Therefore, we next focus on improving the estimation process. Increasing the level of hierarchy in our inference procedure permits the systematic removal of redundant parameters. This is achieved through the hierarchical Bayesian framework and NSBL, \textcolor{comments}{as described next}. 


For NSBL, the inference problem can be stated as in Section \ref{sec:2.3.2}. Although the NSBL algorithm is equipped with the inclusion of hybrid priors for generality, all parameters are assumed to be unknown \textcolor{comments}{in this investigation}. Therefore, ARD priors were assigned to all weight and bias parameters. Considering the overparameterized \textcolor{mine}{(1-3-1)} network, the ARD prior is defined as follows

\begin{equation} \label{eq_4}
\begin{aligned}
\text{p}(\mathbf{\phi}_\alpha \vert \alpha) =  &\mathcal{N}(W^{[1]}_{11} \vert 0,\alpha_{W^{[1]}_{11} }^{-1}) \mathcal{N}(W^{[1]}_{21}  \vert 0,\alpha_{W^{[1]}_{21}}^{-1}) \mathcal{N}(W^{[1]}_{31} \vert 
0,\alpha_{W^{[1]}_{31} }^{-1})\\
&\mathcal{N}(b_1^{[1]}  \vert 0,\alpha_{b^{[1]}_{1}}^{-1})\mathcal{N}(b^{[1]}_{2} \vert 0,\alpha_{b^{[1]}_{2}}^{-1}) \mathcal{N}(b^{[1]}_{3}  \vert 0,\alpha_{b^{[1]}_{3}}^{-1})\\
&\mathcal{N}(W^{[2]}_{11} \vert 0,\alpha_{W^{[2]}_{11} }^{-1}) \mathcal{N}(W^{[2]}_{12}  \vert 0,\alpha_{W^{[2]}_{12}}^{-1})\\
&\mathcal{N}(W^{[2]}_{13} \vert 0,\alpha_{W^{[2]}_{13} }^{-1})\mathcal{N}(b^{[2]}_{1}  \vert 0,\alpha_{b^{[2]}_{1}}^{-1}).
\end{aligned}
\end{equation}

\begin{figure*}[!t]
\centering
\subfloat{\includegraphics[width=2.0in]{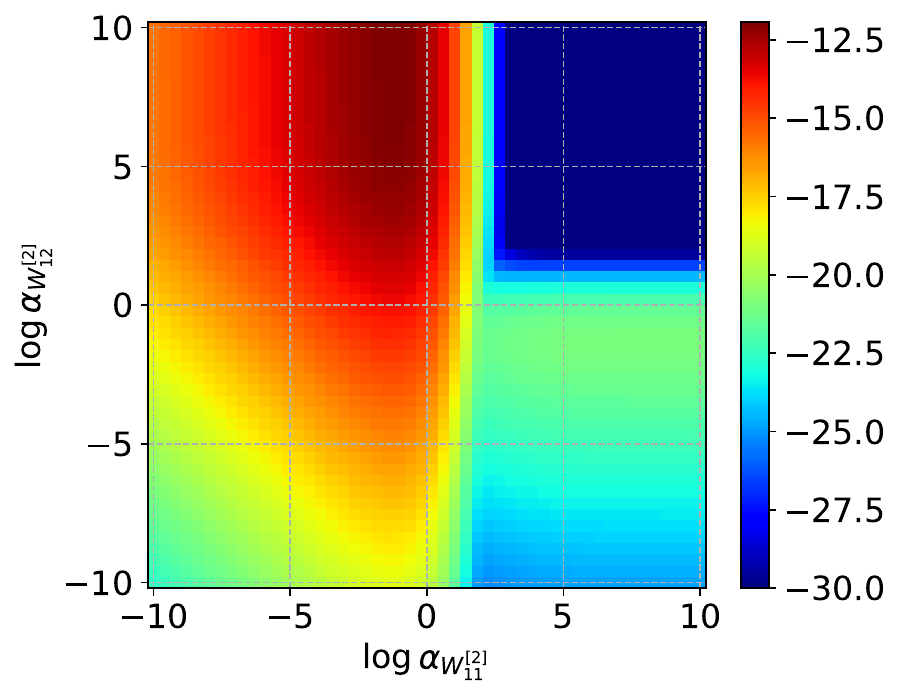}%
}
\hfil
\subfloat{\includegraphics[width=2.0in]{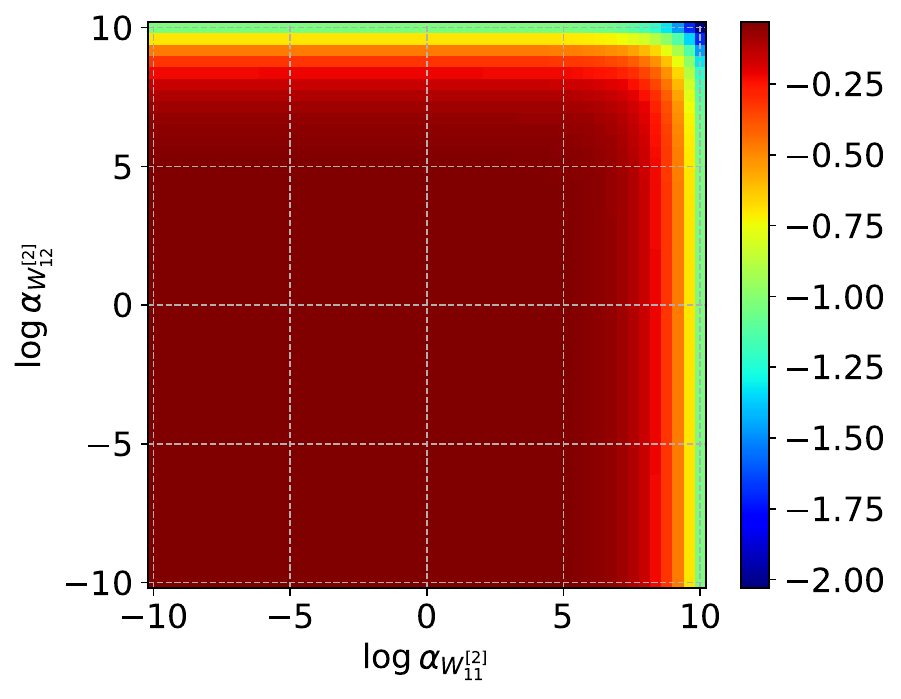}%
}
\hfil
\subfloat{\includegraphics[width=2.0in]{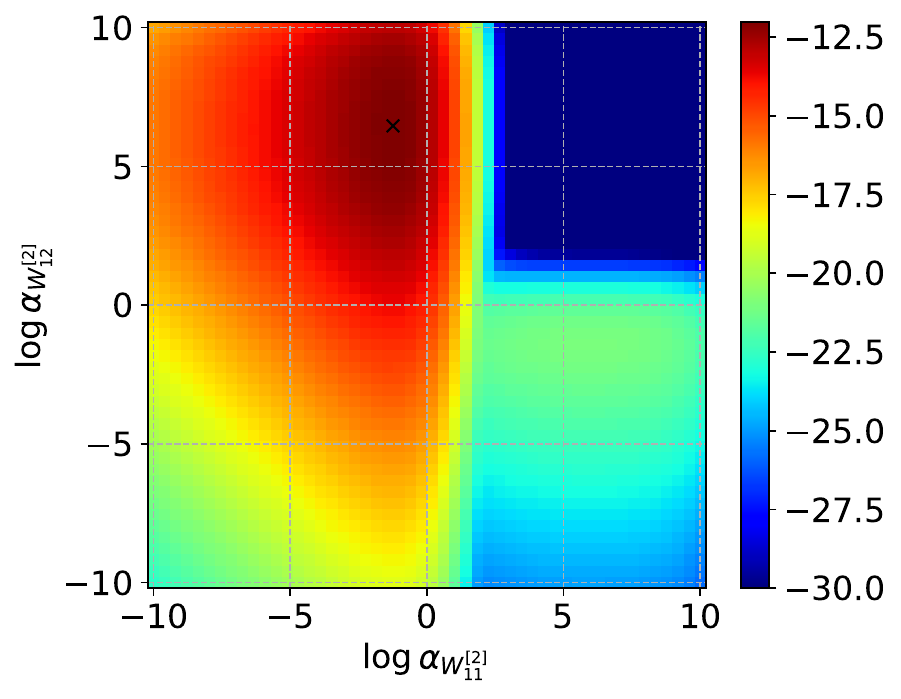}%
}
\hfil\\
\subfloat{\includegraphics[width=2.0in]{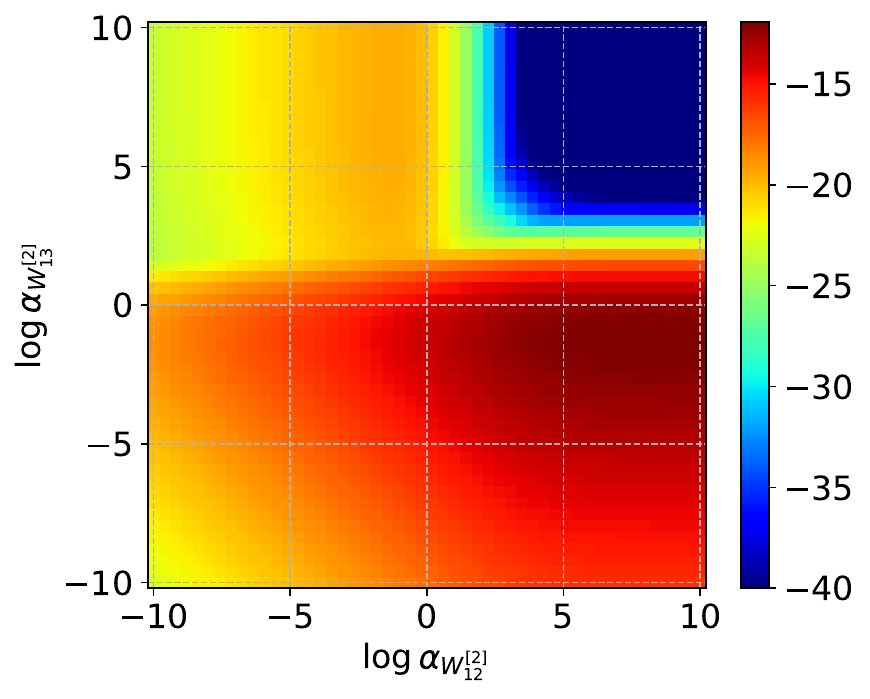}%
}
\hfil
\subfloat{\includegraphics[width=2.0in]{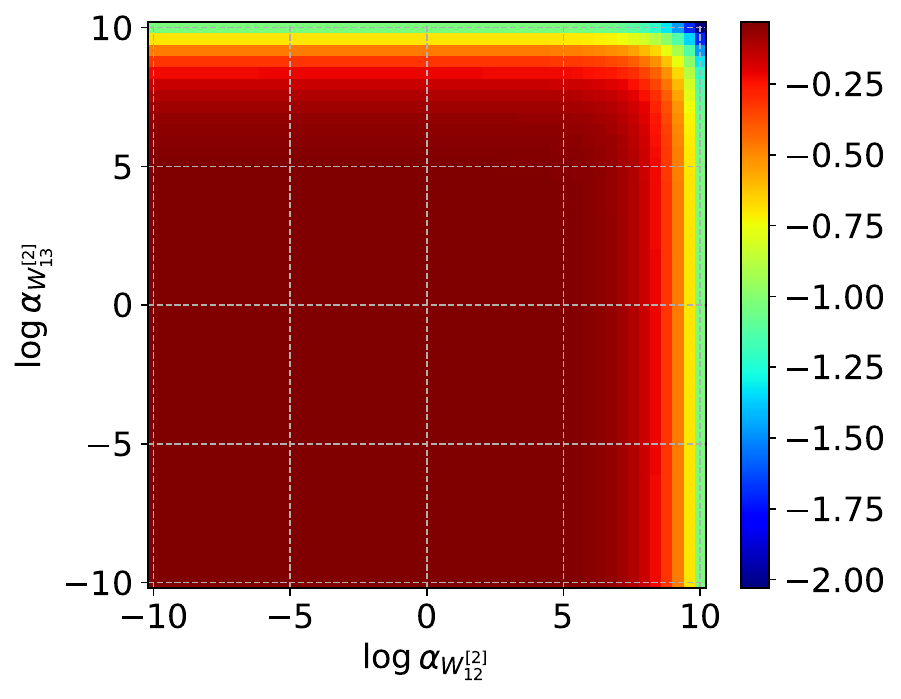}%
}
\hfil
\subfloat{\includegraphics[width=2.0in]{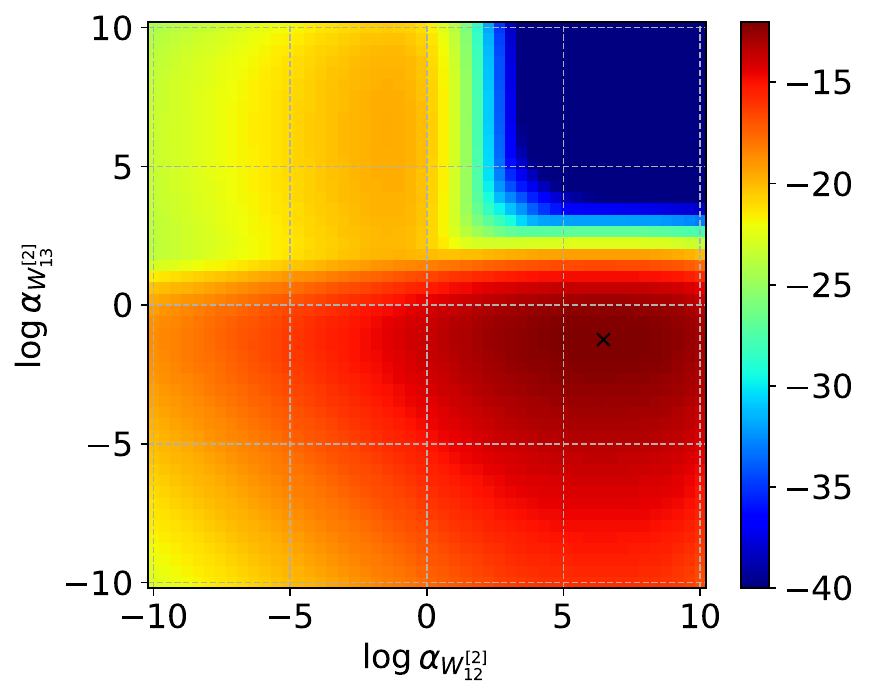}%
}
\hfil\\
\subfloat{\includegraphics[width=2.0in]{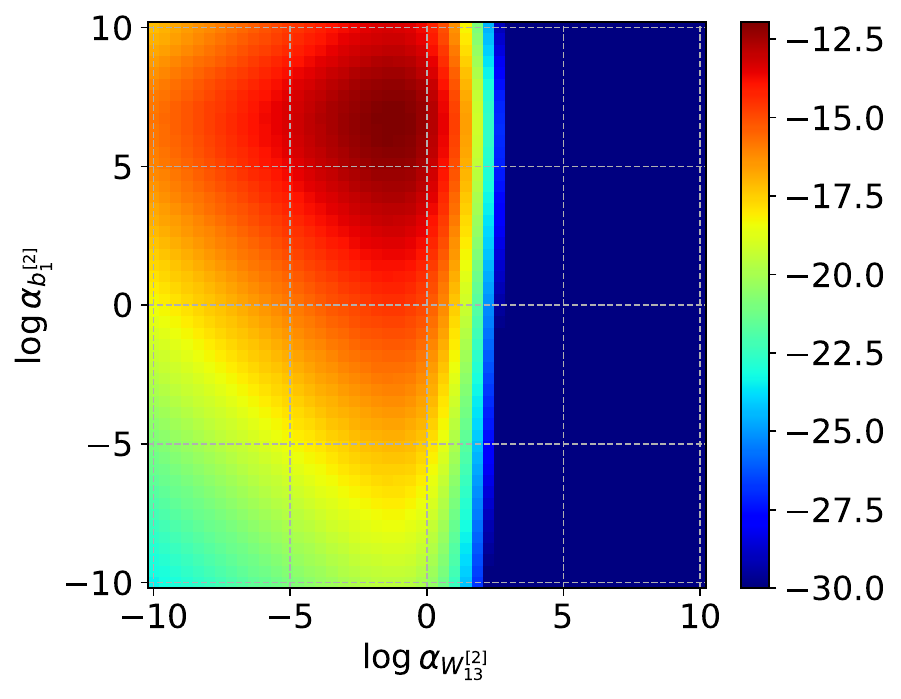}%
}
\hfil
\subfloat{\includegraphics[width=2.0in]{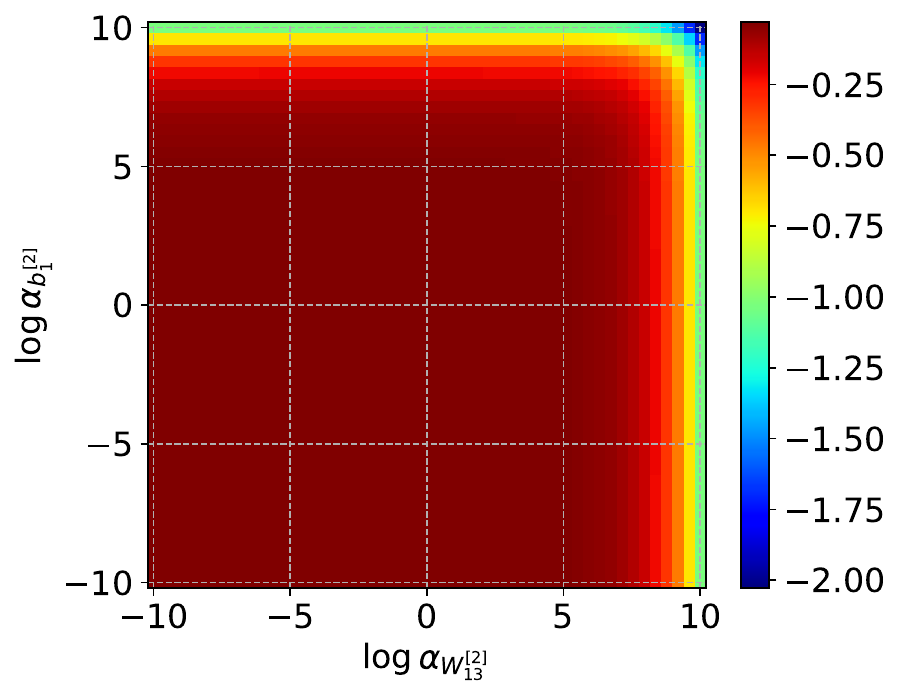}%
}
\hfil
\subfloat{\includegraphics[width=2.0in]{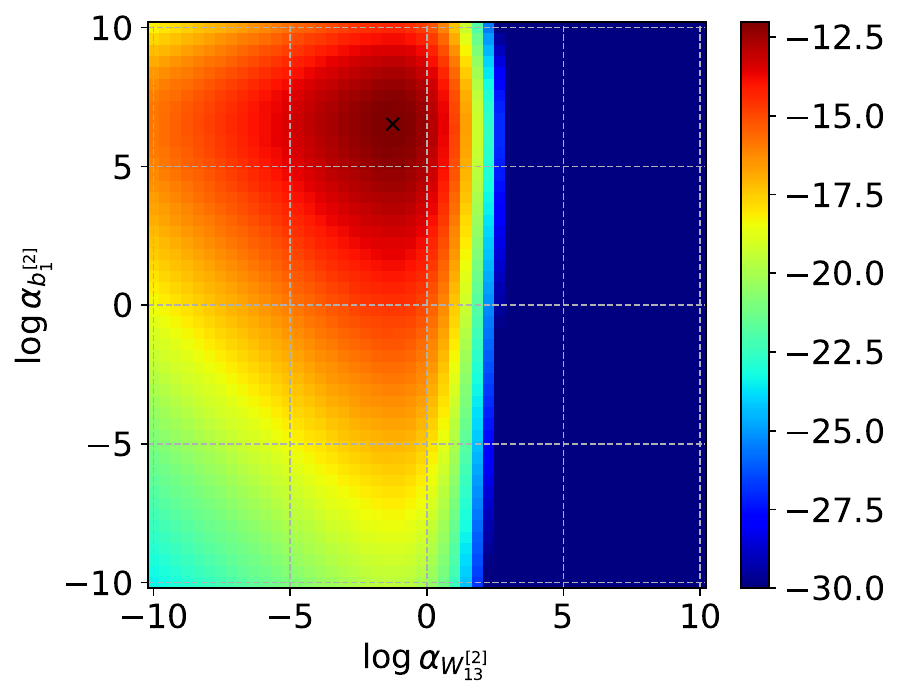}%
}
\caption{Three-dimensional surface plots of the pairwise joint parameter posterior pdfs of the weights between the hidden layer and the output layer; \textcolor{mine}{(First column: evidence, second column: hyperprior, third column: objective function)}}
\label{fig:surfplots}
\end{figure*}

Commonly, adding one level of hierarchy, we are concerned with the probability distribution of the hyperparameters $\mathrm{p}(\boldsymbol{\alpha})$. However, NSBL seeks the MAP estimation of $\boldsymbol{\alpha}$ and not its posterior pdf. Hence, the sampling \textcolor{comments}{is required only for} the product of the likelihood and the non-informative priors, as needed for the construction of the GMM in Eq. (\ref{2:GMM}). As the sampling process does not rely on $\boldsymbol{\alpha}$, it is performed only once. Using the semi-analytical approximations that are conveniently available through the NSBL framework, we can directly obtain estimates of the model evidence and objective function as a function of hyperparameters $\boldsymbol{\alpha}$ \textcolor{comments}{(see Section \ref{sec:2.3.2})}. 
To illustrate the effect of the flat hyperparameter prior, Figure~\ref{fig:surfplots} plots the estimate of the model evidence, the hyperprior, and the resulting objective function \textcolor{comments}{ for hyperparameter pairs while setting other hyperparameters fixed at the MAP estimate. For brevity}, we have chosen to plot a carefully selected subset of parameters in the second layer. These plots serve as representative examples and offer a clear understanding of the impact of the flat hyperprior. In \textcolor{mine}{the second column of} Figure~\ref{fig:surfplots}, the hyperprior \textcolor{comments}{(assumed to be the same for all hyperparameters)} is structured in a way that ensures $\log \text{p}(\log \bm{\alpha})$ remains relatively constant across a significant portion of the range as plotted. However, it decreases near the upper boundaries. This characteristic effectively regulates the objective function, leading to a well-defined optimum shown \textcolor{mine}{by $\times$,} in the last column of Figure~\ref{fig:surfplots}.

\begin{figure*}[!t]
\centering
\subfloat[]{\includegraphics[width=1.6in]{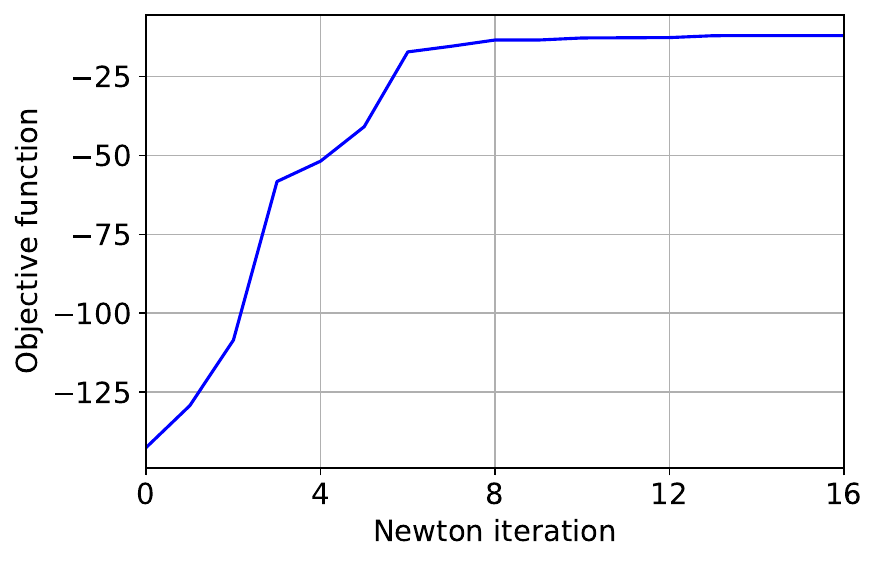}%
\label{fig:plot_objfun_a}}
\hfil
\subfloat[]{\includegraphics[width=1.6in]{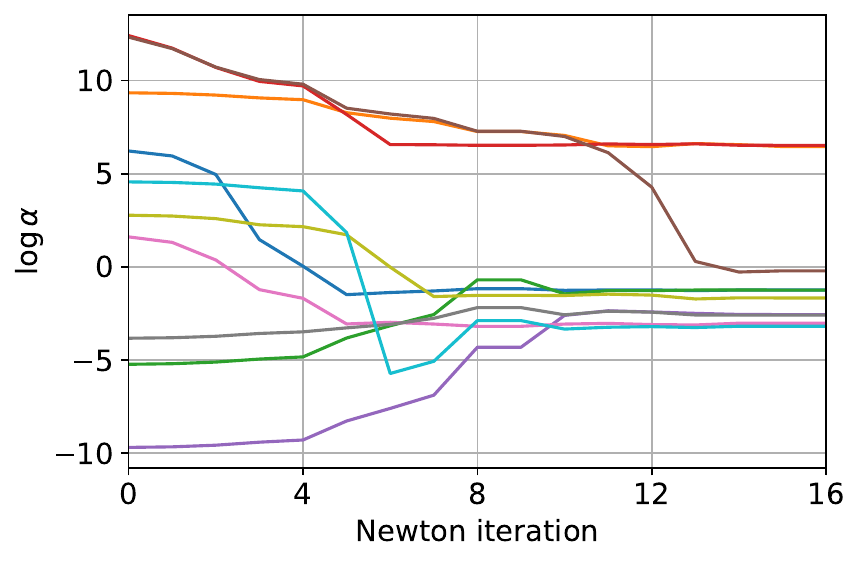}%
\label{fig:plot_logalpha_a}}
\hfil
\subfloat[]{\includegraphics[width=1.6in]{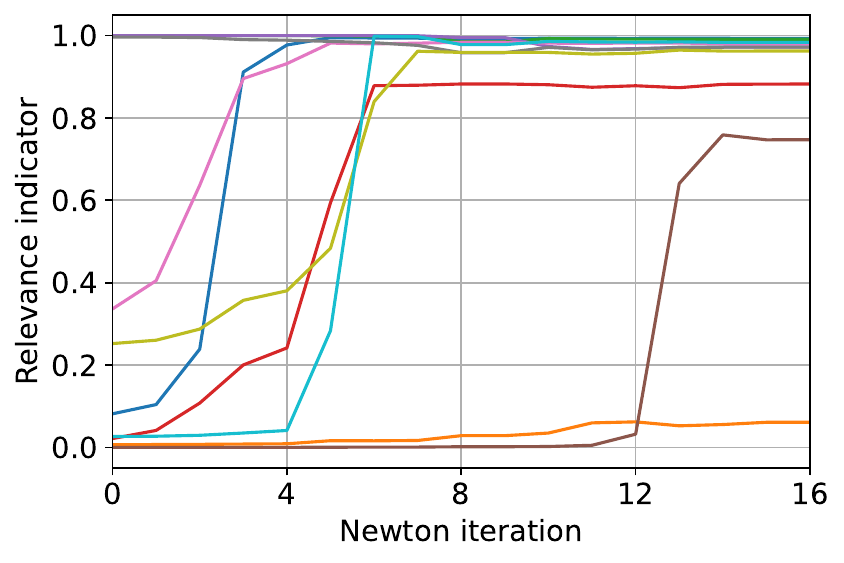}%
\label{fig:plot_rel_a}}
\hfil
\subfloat{\includegraphics[width=0.35in]{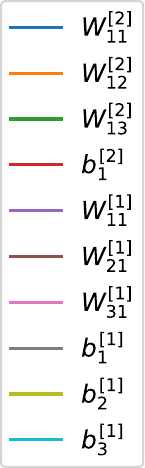}%
}
\caption{NSBL optimization as a function of Newton iterations.}
\label{fig:nsbl_optimization-a}
\end{figure*}

Figure~\ref{fig:plot_objfun_a} \textcolor{comments}{plots values of objective function against} Newton's iterations which \textcolor{mc}{terminates based on the chosen stopping criteria} in 16 iterations. A multi-start \textcolor{comments}{Newton's method} is used \textcolor{comments}{in an effort to discover possible} multiple local optima (if they exist). For brevity, we simply show the results for the \textcolor{comments}{dominant} optimum \textcolor{comments}{discovered through the multistart method}. As we obtain semi-analytical expressions for the objective function (in terms of the kernels of the GMM-based approximation), similar expressions for its gradient vector and Hessian matrix are also available \cite{sandhu2020model,sandhu2021}. Using a Newton trust-region approach \cite{conn2000trust} for optimization \textcolor{mine}{(which removes
the restriction of positive definiteness of the Hessian matrix  making it applicable to non-convex  problems  such as Eq.~\ref{2:logalphaMAP})}, the convergence to the optimal values of the hyperparameters is achieved in relatively few iterations in Figure~\ref{fig:plot_logalpha_a}.  The relevance indicator for all parameters except the subset of $\{W^{[1]}_{21}, b_1^{[2]}, W^{[2]}_{12}\}$ converge close to one, \textcolor{mine}{ as shown in Figure~\ref{fig:plot_rel_a} and  summarized in Table \ref{table:map}. Parameters $\{W^{[1]}_{21}, b_1^{[2]}\}$ have $\gamma_i^{rms}$ values of 0.747 and 0.882, which are inconclusive, and the determination of relevance based on these values is up to the modelers' discretion. Whereas the relevance indicator of 0.061 for parameter $W^{[2]}_{12}$ critically implies irrelevance. Conversely, the high relevance indicator for the other parameters confirms their relevance and yields the structure of the optimal nested network. Though parameters $W_{21}^{[1]}$ and $b_2^{[1]}$ are not necessarily recognized as redundant parameters, the desired outcome of removing the contribution of this neuron is still accomplished by setting $W_{12}^{[2]}$ to zero. Referring to Eq.~(\ref{eq-overparam}), it is evident that when the parameters $W_{12}^{[2]}$ is equal to zero, the entire term associated with the second neuron, $W^{[2]}_{12} \tanh(W^{[1]}_{21} \mathrm{x} + b_2^{[1]})=0$.}
Moreover, due to the aforementioned symmetry of the network, one should expect similar optima corresponding to \textcolor{comments}{two other} cases where either $\{W^{[1]}_{11}, b_1^{[2]}, W^{[2]}_{11}\}$ or $\{W^{[1]}_{31}, b_1^{[2]}, W^{[2]}_{13}\}$ are irrelevant.
The converged values of $\log \alpha_i^{\text{MAP}}$ and the relevance indicator $\gamma_i^{\text{rms}}$ are summarized in Table \ref{table:map}

\begin{figure*}[ht!]
\begin{center}
\includegraphics[scale=0.02]{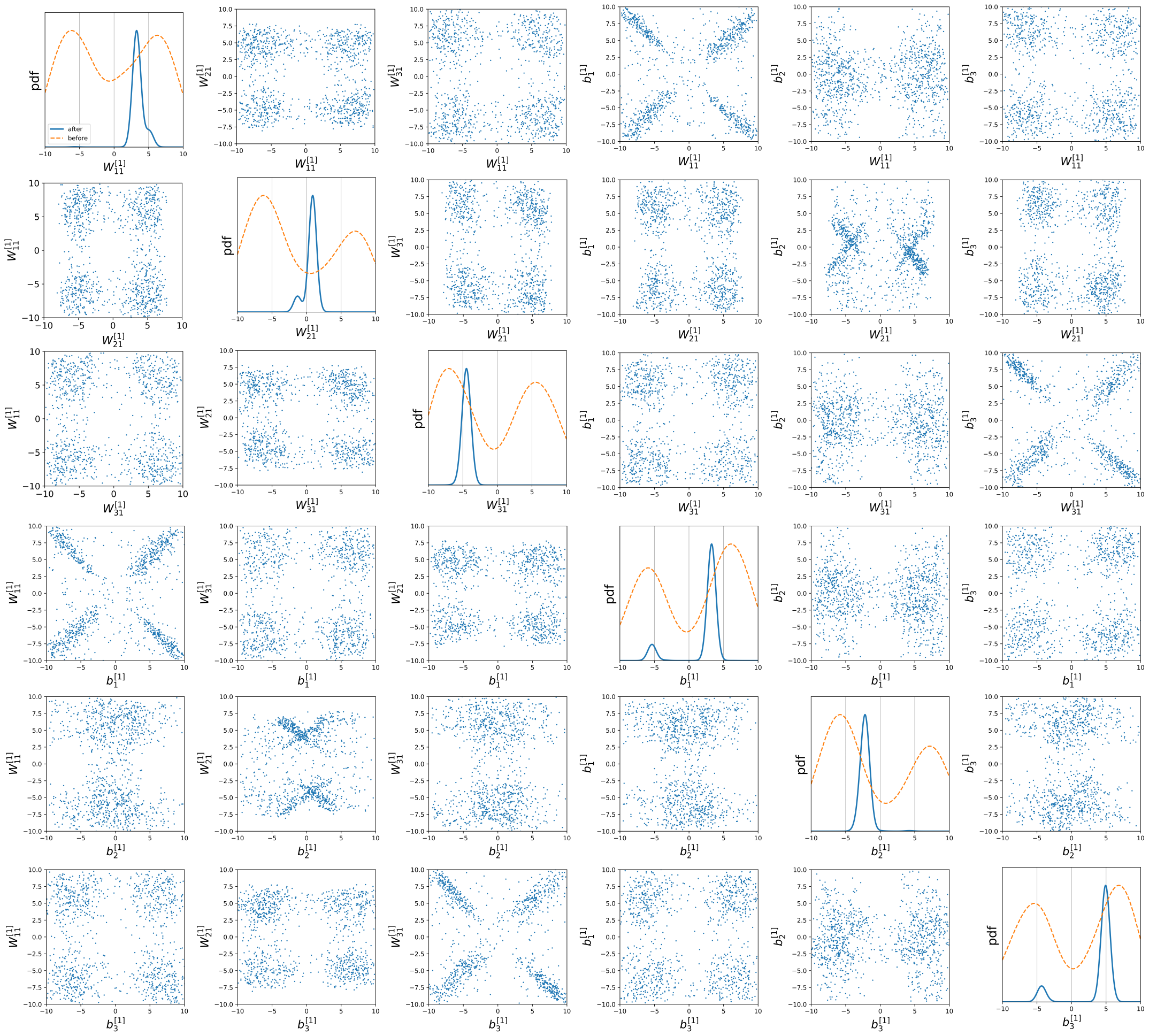}
\end{center}
\caption{Marginal and two-dimensional (pairwise) joint posterior pdfs of weight and bias parameters from the input layer to the hidden layer of the 1-3-1 NN  obtained using NSBL. The
label before (dashed curve) indicates the results obtained by standard Bayesian inference with non-informative priors; this is equivalent to the NSBL results before incorporating the effect of hyperparameters. The label after (solid curve) indicates the results after optimizing the hyperparameters using NSBL.}
\label{mpdf-NSBL-1}
\end{figure*}

\begin{figure}[ht!]
\begin{center}
\includegraphics[scale=0.02]{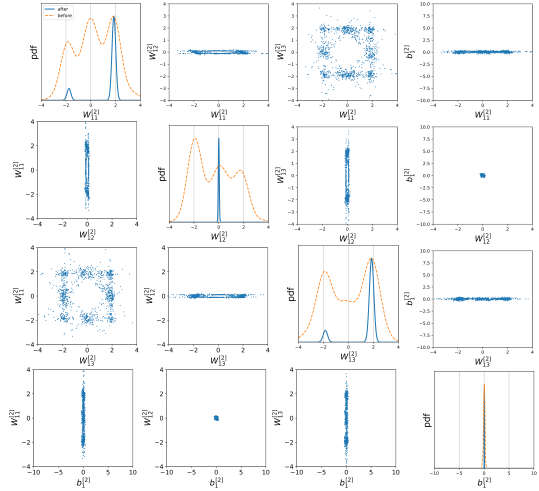}
\end{center}
\caption{Marginal and two-dimensional (pairwise) joint posterior pdfs of  weight and bias parameters from the hidden layer to the output layer of the 1-3-1 NN  obtained using NSBL. The
label before (dashed curve) indicates the results obtained by standard Bayesian inference with non-informative priors; this is equivalent to the NSBL results before incorporating the effect of hyperparameters. The label after (solid curve) indicates the results after optimizing the hyperparameters using NSBL.}
\label{mpdf-NSBL-2}
\end{figure}

In Figure \ref{mpdf-NSBL-1} and \ref{mpdf-NSBL-2}, it is noticeable posterior distributions are more precise after sparse learning, and \textcolor{comments}{other relevant} parameters are captured by the NSBL algorithm \textcolor{comments}{(also confirmed in Table \ref{table:map})}. The presence of multiple peaks in the posterior pdfs indicates that certain parameters exhibit multimodality. This \textcolor{comments}{fact} highlights the possibility of obtaining incorrect uncertainty estimates \textcolor{comments}{in the prediction} when relying solely on the mode of the posterior pdfs. 

\begin{table}[!h]
\centering
\caption{MAP of the hyperparameters and the relevance indicator associated with the network weights and biases}
\begin{tabular}{rrr|rrr}
\hline
\multicolumn{3}{c}{$\Big.$Input layer to hidden layer} & \multicolumn{3}{c}{Hidden layer to output layer} \\
\hline
$\Big.$ Parameter & $\log \alpha_i^{\text{MAP}}$ & $\gamma_i^{\text{rms}}$ & $\Big.$ Parameter & $\log \alpha_i^{\text{MAP}}$ & $\gamma_i^{\text{rms}}$\\
\hline		
$\Big. W^{[1]}_{11}$ & -2.560 & 0.972& $\Big. W^{[2]}_{11}$ & -1.237 & 0.993\\
$\Big. W^{[1]}_{21}$ & -0.211 & 0.747 & $\Big. W^{[2]}_{12}$ &6.460 &0.061 \\
$\Big. W^{[1]}_{31}$ & -3.025 & 0.980 & $\Big. W^{[2]}_{13}$ & -1.246 & 0.991\\
$\Big. b^{[1]}_{1}$ & -2.589 & 0.972& $\Big. b^{[2]}_{1}$ & 6.524 & 0.882 \\
$\Big. b^{[1]}_{2}$ & -1.667 & 0.962 & & &\\
$\Big. b^{[1]}_{3}$ & -3.182 & 0.984 & & &\\
\hline \hline		
\end{tabular}

\label{table:map}
\end{table}

\subsubsection{\bf NSBL with the Laplace approximation}
In Figures \ref{fig:2neurons_standard_post}, \ref{fig:jpdf_Bayes-lay1} and \ref{fig:jpdf_Bayes-lay2}, we have illustrated the multimodality in the parameter posterior pdfs, resulting from the inherent symmetry in the NN. In addition to the multimodality, the pairwise correlation structure of the parameter pdfs is also evident through these plots. For a comparison, we perform a Laplace approximation \textcolor{comments}{whereby the likelihood is approximated to be a Gaussian \cite{murphy2012, bishop2006}}. \textcolor{comments}{When} a Laplace approximation of the likelihood is used in conjunction with Gaussian ARD priors for all parameters, it effectively reduces the NSBL algorithm to SBL/RVM. 

In the example of the 1-3-1 network, we observed that the parameter posterior was highly multimodal and thus clearly non-Gaussian. However, since the multi-modality was a direct result of the symmetry of the network, a Laplace approximation centered at one of these modes may \textcolor{comments}{provide reasonable} results insofar as the discovery of the optimal nested model structure is concerned.

\begin{figure*}[ht!]
\begin{center}
\includegraphics[scale=0.2]{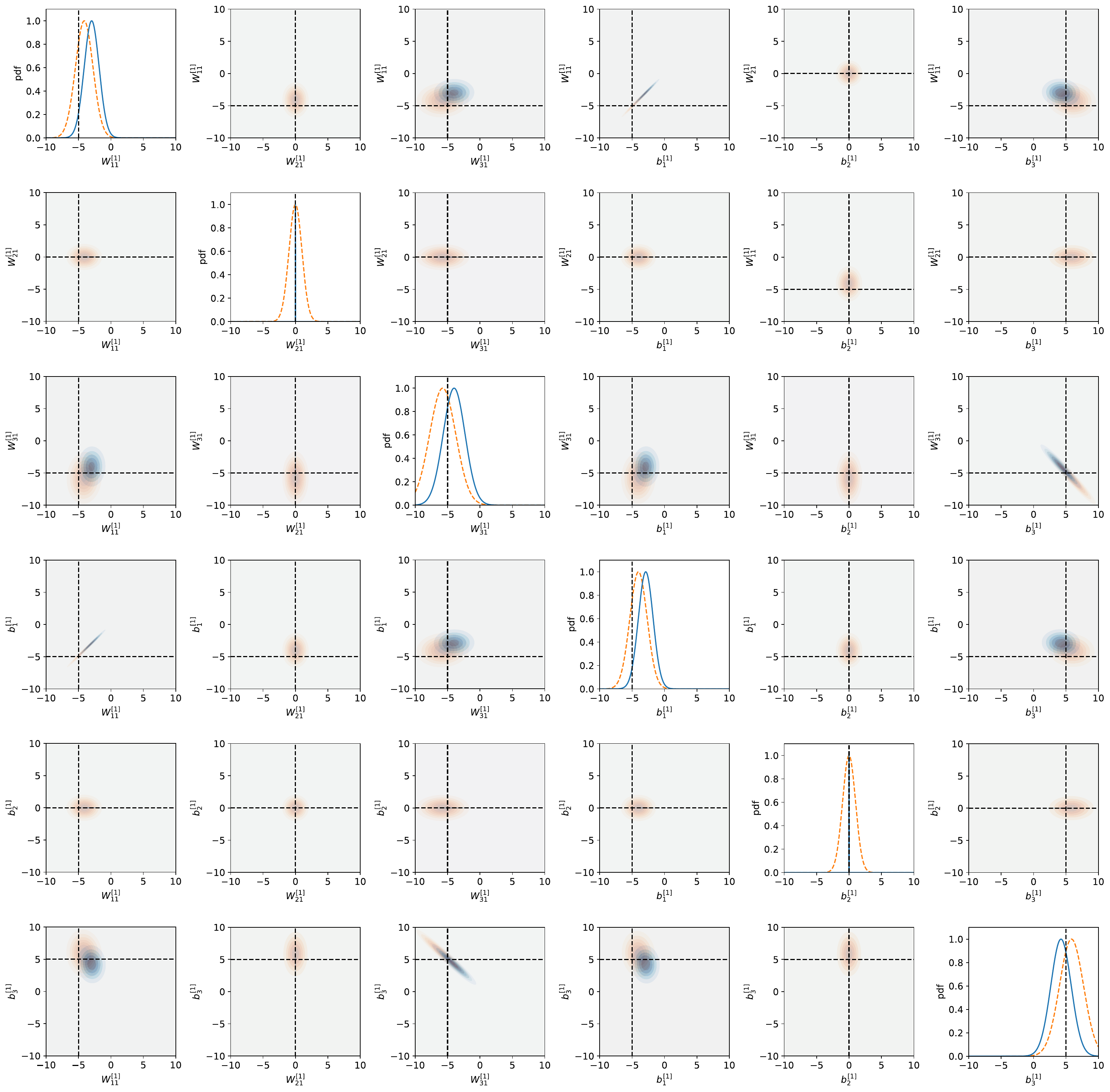}
\end{center}
\caption{Laplace approximation with and without NSBL: Marginal and joint posterior pdf of weight and bias parameters between the input layer and the hidden layer before and after sparsifying}
\label{mpdf-laplace-1}
\end{figure*}

\begin{figure}[ht!]
\begin{center}
\includegraphics[scale=0.2]{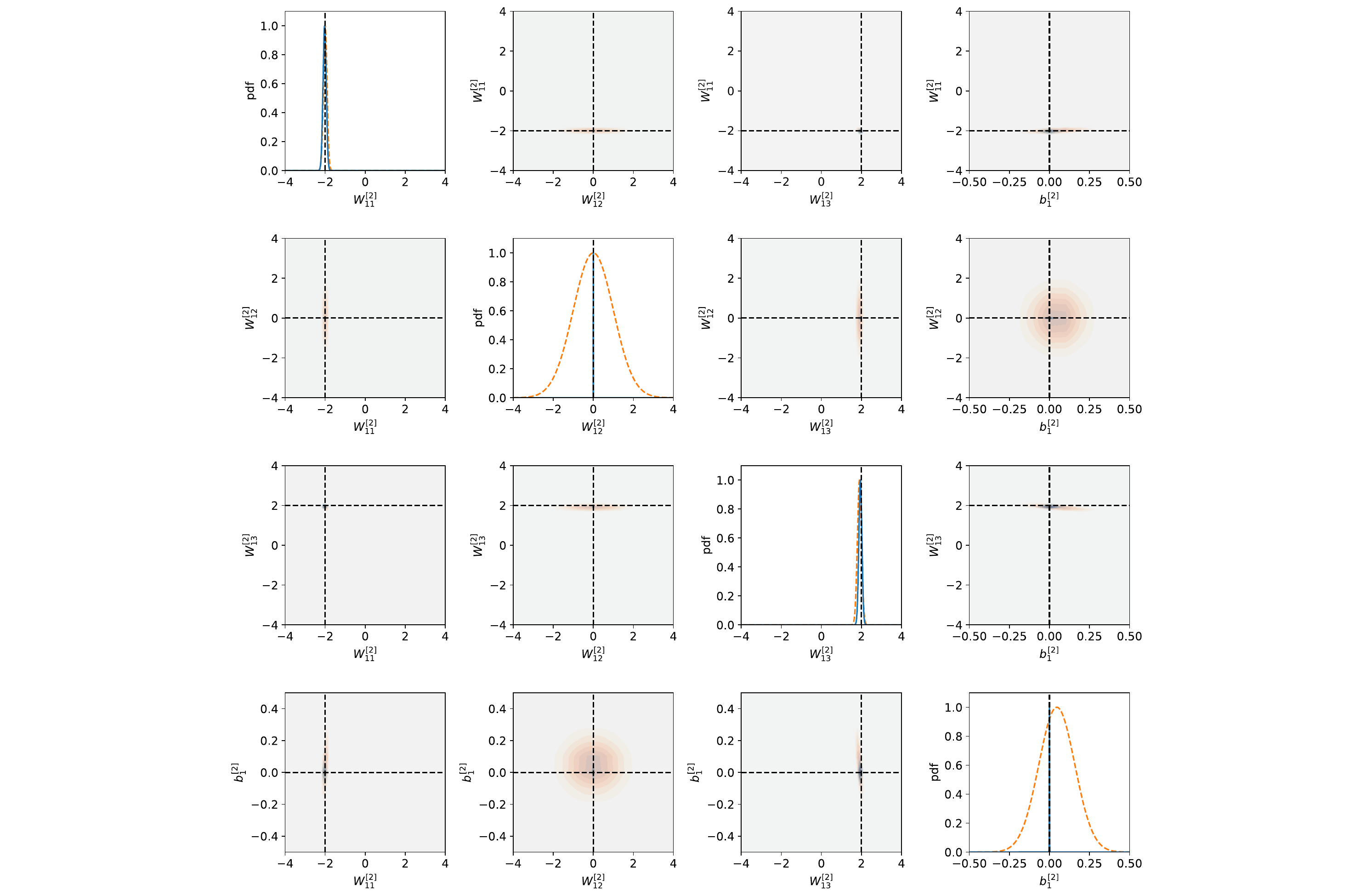}
\end{center}
\caption{Laplace approximation with and without NSBL: Marginal and joint posterior pdf of weight and bias parameters between the hidden layer and the output layer before and after sparsifying}
\label{mpdf-laplace-2}
\end{figure}

The Laplace approximation is constructed at one of the eight optima corresponding to the combination where $W_{12}^{[2]}$, $W_{21}^{[1]}$, and $b_2^{[1]}$ are zero. The MAP estimate of the parameters and the diagonal elements of the covariance matrix are presented in the leftmost columns of Table \ref{table:map_la}. The important correlation between the parameters observed from the TMCMC samples in Figure \ref{fig:3d_mpost} is captured in the off-diagonal elements of the covariance matrix, however they are not reported (nor are the joint pdfs shown) for brevity. The rightmost columns of Table \ref{table:map_la} summarize the sparse learning results, presenting the optimized log of the hyperparameters \textcolor{comments}{denoted by $\log \alpha_i^{\text{MAP}}$}, the corresponding relevance indicators, and the shifted mean and variance after sparse learning. The marginal and pairwise joint posterior pdfs before and after sparse learning are presented in Figures \ref{mpdf-laplace-1} and \ref{mpdf-laplace-2}. 

\begin{table}[!h]
\centering
\caption{MAP and the covariance matrix (diagonal entries) of the Laplace approximation before and after sparse learning. The MAP of the hyperparameters and relevance indicator associated with the network weights and biases are also included.}
\begin{tabular}{rrr|rrrr}
\hline
\multicolumn{3}{c}{$\Big.$Laplace approximation} & \multicolumn{3}{c}{Sparse learning} \\
\hline
$\Big.$ Parameter & $\phi_i^{\text{MAP}}$ & $\Sigma_{ii}$ & $\Big.$ $\log \alpha_i^{\text{MAP}}$ & $\gamma_i^{\text{rms}}$ & $\textbf{m}_{i}$ & $\textbf{P}_{ii}$\\
\hline		
$\Big. W^{[1]}_{11}$ & -1.994 & 0.008 & -1.416 & 0.999& -2.028 & 0.006\\
$\Big. W^{[1]}_{21}$ & 0.000 & 1.000 & 12.153 & 0.000 & 0.000 & 0.000\\
$\Big. W^{[1]}_{31}$ & 1.903 & 0.011 & -1.335 &0.999 & 1.948 & 0.006\\
$\Big. b^{[1]}_{1}$ & 0.046 & 0.013 & 14.327 & 0.000& 0.000 & 0.000\\
$\Big. b^{[1]}_{2}$ & -4.121 & 1.826 &  -2.32 & 0.870 & -2.974 & 1.326\\
$\Big. b^{[1]}_{3}$ & 0.000 & 1.000 & 12.153 & 0.000 & 0.000 & 0.000\\
 \hline
$\Big.  W^{[2]}_{11}$ &  -5.753 & 4.077 & -2.934& 0.848 & -3.993 & 2.862\\
$\Big. W^{[2]}_{12}$ & -4.016 & 1.769 & -2.270 & 0.867 & -2.896 & 1.291\\
$\Big. W^{[2]}_{13}$ & 0.000 & 1.000 &12.153 & 0.000 & 0.000 & 0.000 \\
$\Big. b^{[2]}_{1}$ & 5.849 & 3.411 &-3.014 & 0.880 & 4.233 &2.455 \\
\hline \hline		
\end{tabular}

\label{table:map_la}
\end{table}

\textcolor{mine}{Note that in Table \ref{table:map_la}, we report the mean and diagonal entries of the covariance matrix before and after sparse learning to show the improvement that is offered in the posterior estimates. We intentionally omit this information in Table \ref{table:map}, as these quantities are not meaningful given the non-Gaussian characteristics of the multimodal parameter posterior pdfs.}

\subsubsection{\bf Hierachical Bayesian Inference} 
\begin{figure}[ht!]
\begin{center}
\includegraphics[scale=0.02]{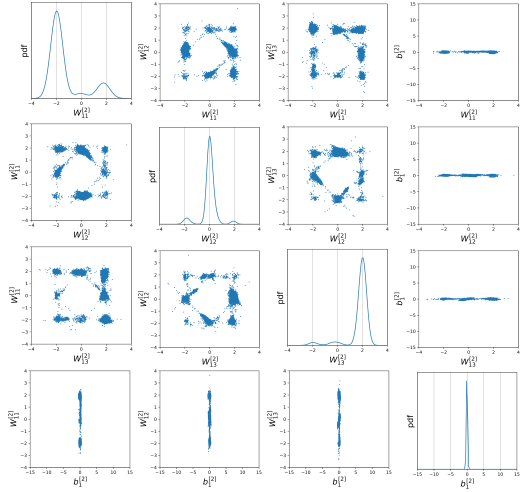}
\end{center}
\caption{Marginal posterior and \textcolor{comments}{two-dimensional} joint posterior of weight and bias parameters from the hidden layer to the output layer of the 1-3-1 NN obtained using hierarchical Bayesian inference.}
\label{fig:jpdf_Hierarchical-lay2}
\end{figure}

\begin{figure*}[ht!]
\begin{center}
\includegraphics[scale=0.02]{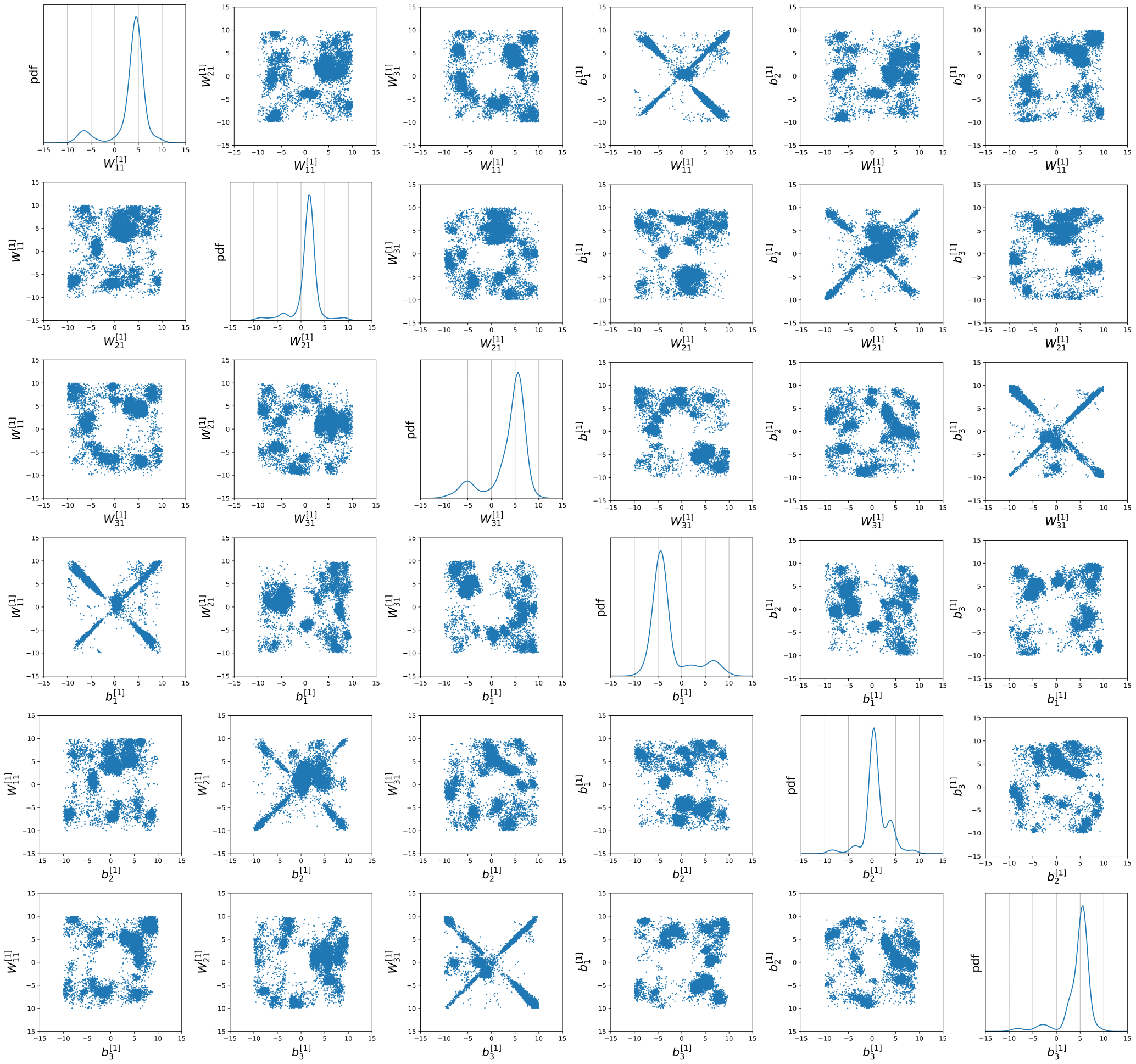}
\end{center}
\caption{Marginal posterior and \textcolor{comments}{two-dimensional} joint posterior of weight and bias parameters from the input layer to the hidden layer of the 1-3-1 NN obtained using hierarchical Bayesian inference.}
\label{fig:jpdf_Hierarchical-lay1}
\end{figure*}

\begin{figure*}[ht!]
\begin{center}
\includegraphics[scale=0.02]{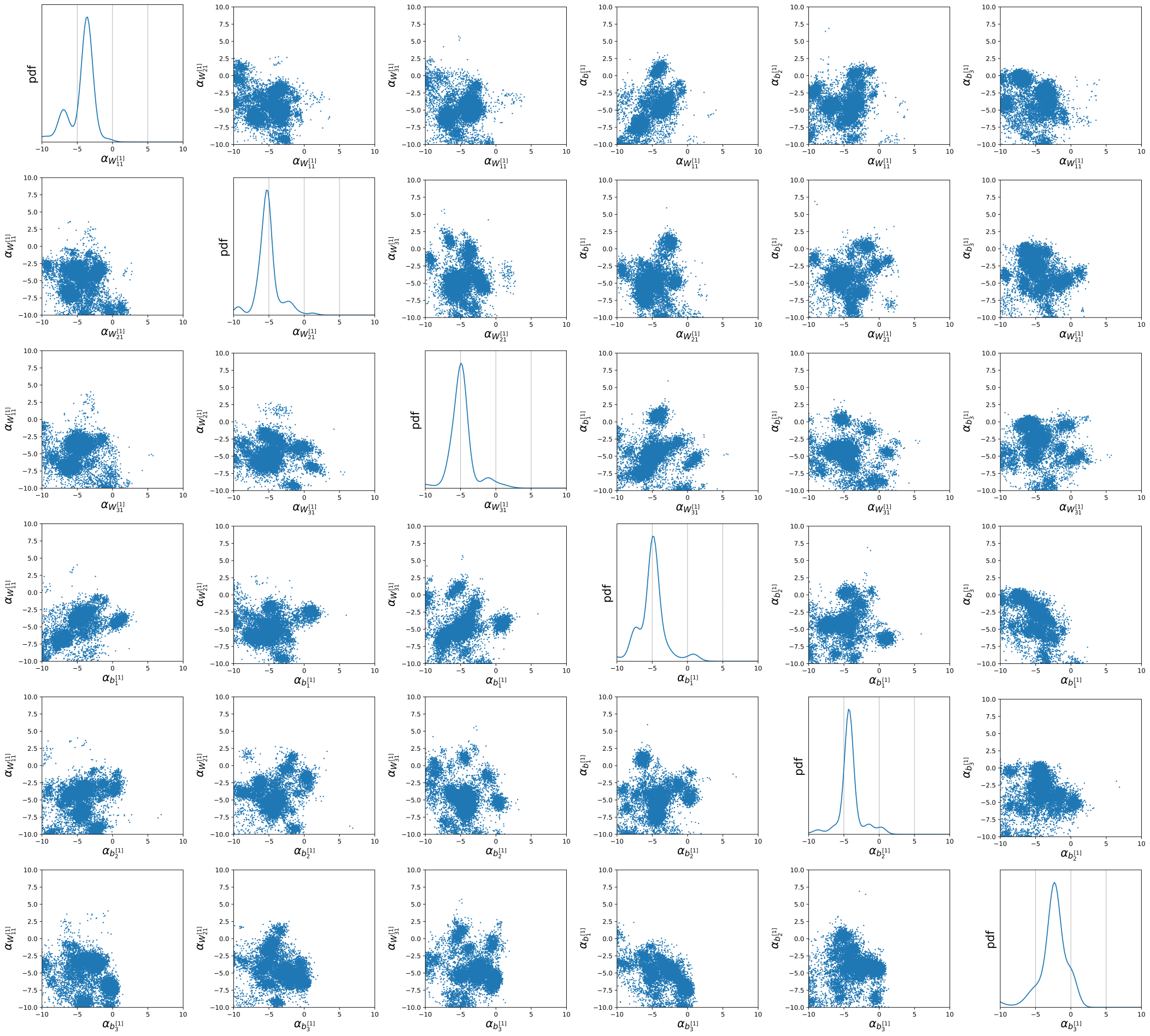}
\end{center}
\caption{Marginal posterior and \textcolor{comments}{two-dimensional} joint posterior of hyperparameters from the input layer to the hidden layer of the 1-3-1 NN obtained using hierarchical Bayesian inference.}
\label{fig:jpdf_Hyperparameter-lay2}
\end{figure*}

\begin{figure*}[ht!]
\begin{center}
\includegraphics[scale=0.02]{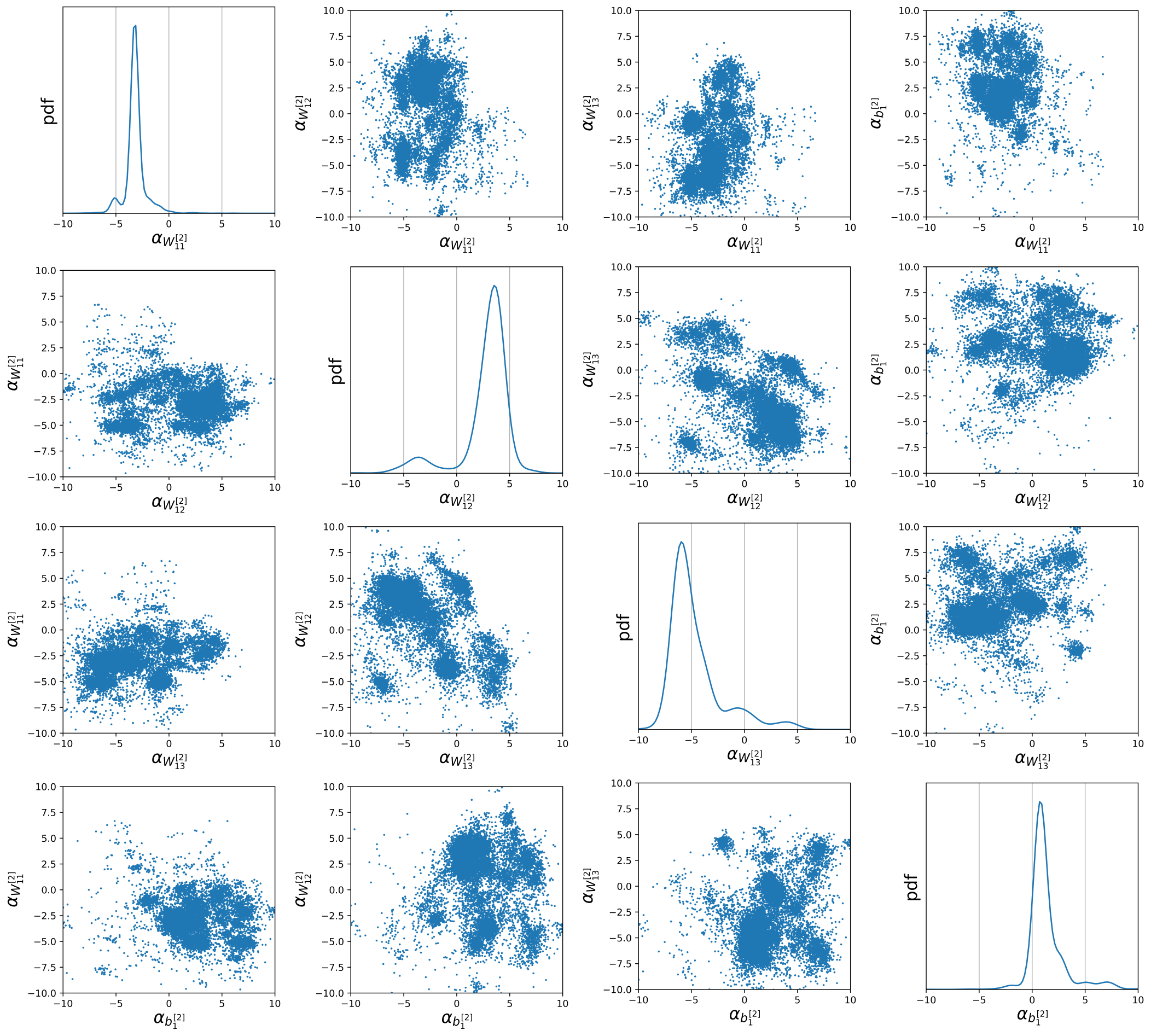}
\end{center}
\caption{Marginal posterior and \textcolor{comments}{two-dimensional} joint posterior of hyperparameters from the hidden layer to the input layer of the 1-3-1 NN obtained using hierarchical Bayesian inference.}
\label{fig:jpdf_Hyperparameter-lay1}
\end{figure*}

While NSBL depends on the MAP estimation of \textcolor{comments}{the hyperparameter} $\boldsymbol{\alpha}$, jointly estimating the parameters $\bm{\phi}$ and the hyperparameters $\boldsymbol{\alpha}$ leads to the hierarchical Bayesian framework as stated as in Eq.~(\ref{2:inference}). Through this setting, all parameters are treated as questionable and are assigned ARD priors. The hyperprior \textcolor{comments}{of the precision $\boldsymbol{\alpha}$ of each ARD parameter prior is assumed to have} a Gamma distribution with shape and rate parameters $r_1 = r_2 = 1+\exp(-10)$ and $s_1 = s_2 = \exp(-10)$. This parameterization of the Gamma hyperprior results in an approximately uniform distribution in the range $\exp(-10) \leq \alpha_1, \alpha_2 \leq \exp(10)$. Therefore, the hyperparameter posterior predominantly \textcolor{comments}{dictated by} data, with the hyperprior setting an upper bound on the precision of a redundant parameter. The resulting parameter posterior pdfs generated through 500,000 samples through the TMCMC algorithm are plotted in Figures \ref{fig:jpdf_Hierarchical-lay2} and \ref{fig:jpdf_Hierarchical-lay1}, displaying distinct multimodality in all parameter posteriors (except $b_1^{[2]})$. \textcolor{comments}{Note the vivid multimodal features in the parameter posterior pdfs in Figure~\ref{fig:jpdf_Hierarchical-lay2} and \ref{fig:jpdf_Hierarchical-lay1} contributed by the non-Gaussian (some also being multimodal) hyperparameter posteriors shown in Figure~\ref{fig:jpdf_Hyperparameter-lay2} and \ref{fig:jpdf_Hyperparameter-lay1}.} 

\subsubsection{\bf Comparison of Standard Bayesian inference, Hierarchical Bayesian inference, NSBL, and Laplace approximation with and without NSBL}
\begin{figure*}[!t]
\centering
\subfloat[]{\includegraphics[width=2.0in]{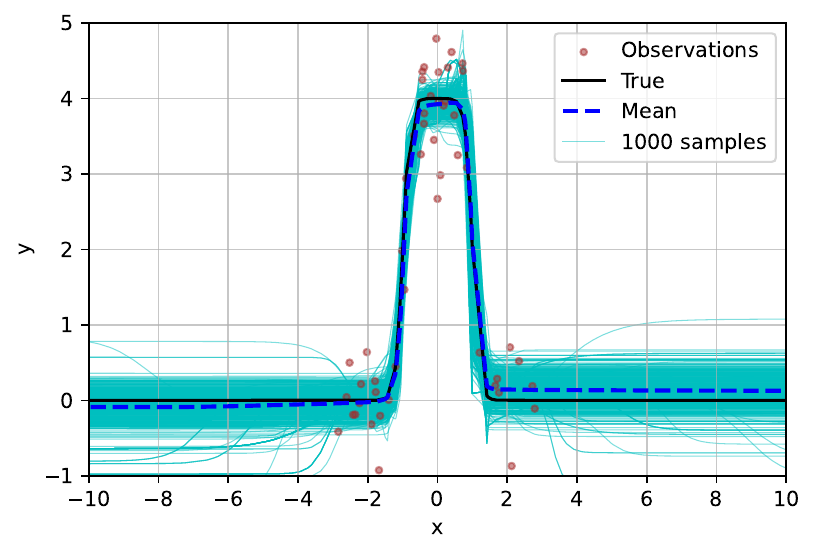}%
}
\hfil
\subfloat[]{\includegraphics[width=2.0in]{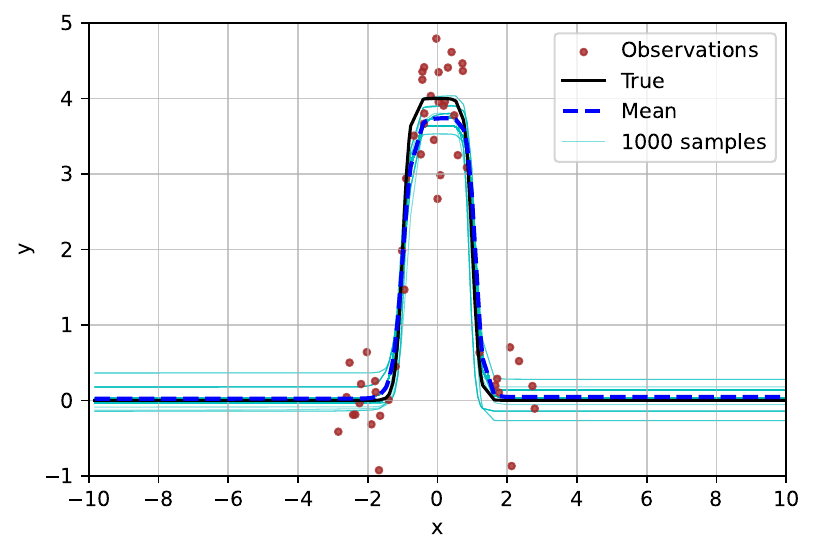}%
}
\hfil
\subfloat[]{\includegraphics[width=2.0in]{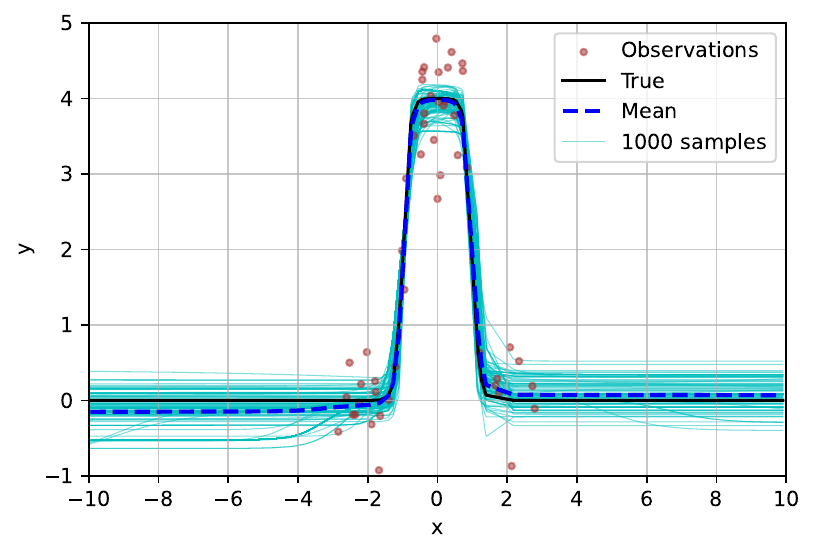}%
}
\hfil
\subfloat[]{\includegraphics[width=2.0in]{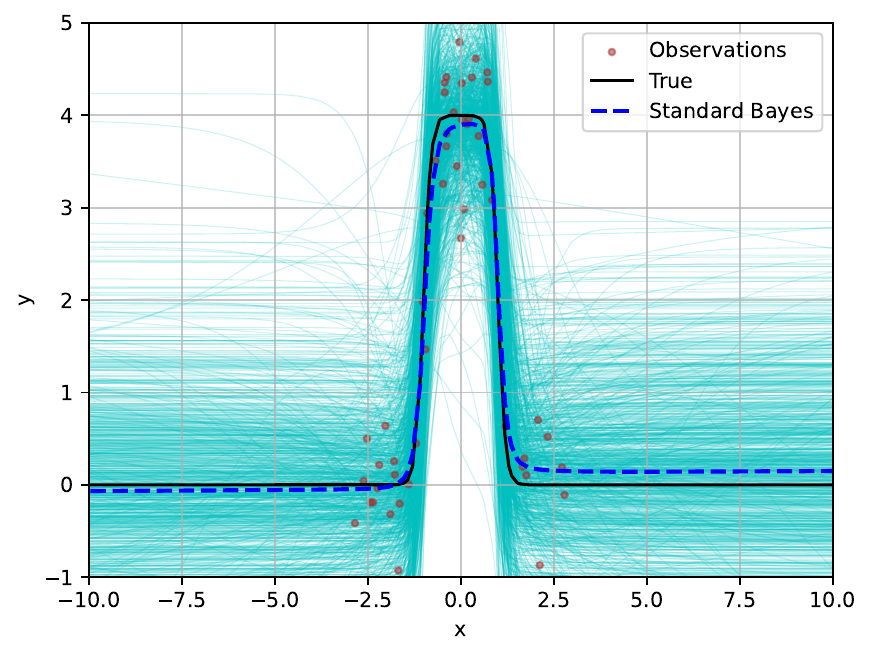}%
}
\hfil
\subfloat[]{\includegraphics[width=2.0in]{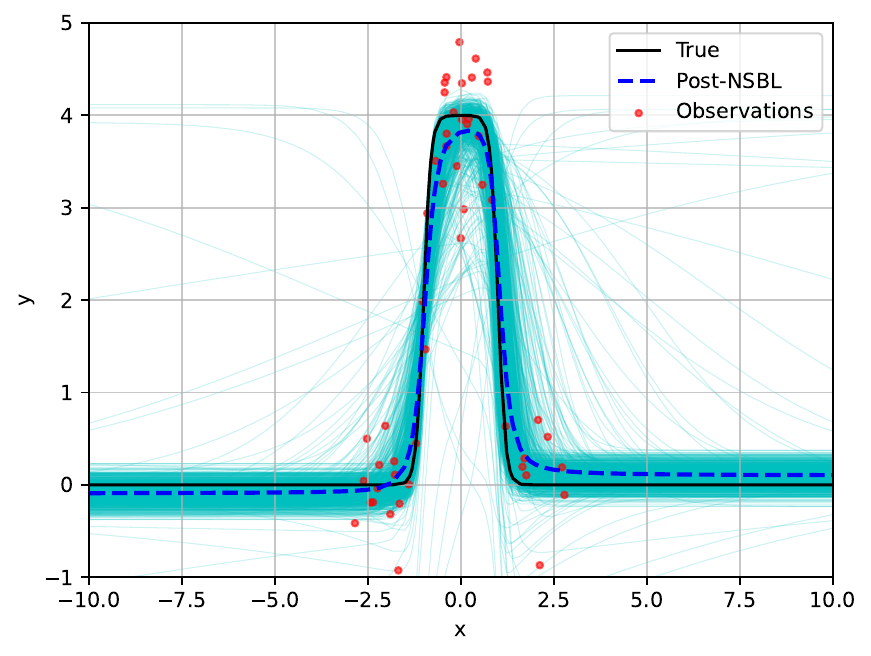}%
}
\caption{1000 samples predictions: (a) Standard Bayesian Inference, (b) NSBL, (c) Hierarchical Bayesian Inference, (d) Laplace approximation, (e) Laplace approximation with NSBL. }
\label{fig:prediction}
\end{figure*}

Figure \ref{fig:prediction} \textcolor{comments}{shows 1000 sample predictions along with the mean prediction using} standard Bayesian inference, NSBL, hierarchical Bayesian inference, and Laplace approximation with and without the  NSBL \textcolor{comments}{for the neural network in Figure~\ref{fig:NN-3}}. \textcolor{mine}{These predictions are constructed based on the samples from the so-called push-forward posterior \cite{butler2018}.}
Considering the similarity in the sparse learning mechanisms of NSBL and hierarchical Bayesian inference, similar sparsity levels were obtained from both algorithms, as evident from Figure~\ref{fig:prediction}-b and \ref{fig:prediction}-c.  

In contrast to the standard Bayesian inference \textcolor{comments}{(Figure~\ref{fig:prediction}-a)} and Laplace approximation with and without NSBL \textcolor{comments}{(Figure~\ref{fig:prediction}-d and \ref{fig:prediction}-e, respectively)}, which suffer from pronounced overfitting, the predictions from hierarchical Bayesian inference \textcolor{comments}{(Figure~\ref{fig:prediction}-c)} and NSBL \textcolor{comments}{(Figure~\ref{fig:prediction}-b)} highlight their effectiveness in \textcolor{comments}{alleviating overfiting problem and consequent reduction in predictive uncertainty.} 

\section{Conclusion}
\label{sec:conclusion}
\textcolor{comments}{This paper reports a computationally efficient Bayesian framework to induce sparsity in over-parameterized NNs in order to alleviate the overfitting issue and consequently reduce uncertainty in the predictions, using the recently proposed NSBL algorithm. The construction of the SBNN proceeds with prescribing sparsity-inducing Gaussian ARD priors for the network parameters (weights and biases). The data-optimal precision parameters (MAP estimates) of the ARD prior are obtained by regularized evidence optimization.
In the current NSBL algorithm, the critical step involves the construction of the Gaussian mixture approximation of the likelihood function using its TMCMC samples. For a regression problem using an overparameterized NN, the benefit of SBNN is highlighted in reducing overfitting and uncertainty in prediction. The comparable predictive performance of the NNs constructed using NSBL and hierarchical Bayesian inference demonstrates the robustness of the NSBL-based SBNN. While both hierarchical Bayesian inference and NSBL rely on TMCMC, the parameter space of sampling for the hierarchical Bayesian inference doubles (vis-a-vis NSBL). This fact points out the computational efficiency of NSBL compared to the hierarchical Bayesian inference. } 

\textcolor{mine}{We have demonstrated the applicability of the SBNN methodology through a shallow NN as an illustrative example. However, extending this to deep NNs involves a dramatic increase in the parameter space. Strategies such as layer-wise learning may permit a practical implementation of the NSBL algorithm for NNs with increasing numbers of parameters.}
\textcolor{comments}{
Furthermore, this research aims to develop a sampling-free approach for GMM construction of the likelihood function, thereby enhancing computational efficiency of the current algorithm.}

\bibliographystyle{unsrt}  
\bibliography{references}  

\begin{thebibliography}{10}

\bibitem{konishi2008information}
Sadanori Konishi and Genshiro Kitagawa.
\newblock {\em Information criteria and statistical modeling}.
\newblock Springer Science \& Business Media, 2008.

\bibitem{murphy2012}
Kevin~P Murphy.
\newblock {\em Machine learning: a probabilistic perspective}.
\newblock MIT press, 2012.

\bibitem{mackay1992}
David~JC MacKay.
\newblock A practical {B}ayesian framework for backpropagation networks.
\newblock {\em Neural computation}, 4(3):448--472, 1992.

\bibitem{neal1992}
Radford~M Neal.
\newblock {B}ayesian training of backpropagation networks by the hybrid {M}onte
  {C}arlo method.
\newblock Technical report, Citeseer, 1992.

\bibitem{mackay1995}
David~JC MacKay.
\newblock Probable networks and plausible predictions - a review of practical
  {B}ayesian methods for supervised neural networks.
\newblock {\em Network: computation in neural systems}, 6(3):469, 1995.

\bibitem{neal1996}
R.~M. Neal.
\newblock {B}ayesian learning for neural networks.
\newblock {\em SpringerVerlag}, 118(448-472), 1996.

\bibitem{izmailov2021}
P.~Izmailov, S.~Vikram, M.~D. Hoffman, and A.~G. Wilson.
\newblock What are {B}ayesian neural network posteriors really like?
\newblock {\em CoRR}, abs/2104.14421, 2021.

\bibitem{jospin2022hands}
Laurent~Valentin Jospin, Hamid Laga, Farid Boussaid, Wray Buntine, and Mohammed
  Bennamoun.
\newblock Hands-on {B}ayesian neural networks—a tutorial for deep learning
  users.
\newblock {\em IEEE Computational Intelligence Magazine}, 17(2):29--48, 2022.

\bibitem{chib1995}
Siddhartha Chib and Edward Greenberg.
\newblock Understanding the {M}etropolis-{H}astings algorithm.
\newblock {\em The american statistician}, 49(4):327--335, 1995.

\bibitem{wilson2020}
A.~G. Wilson and P.~Izmailov.
\newblock {B}ayesian deep learning and a probabilistic perspective of
  generalization.
\newblock {\em CoRR}, abs/2002.08791, 2020.

\bibitem{neal2011}
Radford~M Neal et~al.
\newblock {MCMC} using {H}amiltonian dynamics.
\newblock {\em Handbook of {M}arkov {C}hain {M}onte {C}arlo}, 2(11):2, 2011.

\bibitem{blei2017}
David~M Blei, Alp Kucukelbir, and Jon~D McAuliffe.
\newblock Variational inference: A review for statisticians.
\newblock {\em Journal of the American statistical Association},
  112(518):859--877, 2017.

\bibitem{graves2011practical}
Alex Graves.
\newblock Practical variational inference for neural networks.
\newblock {\em Advances in neural information processing systems}, 24, 2011.

\bibitem{shridhar2019comprehensive}
Kumar Shridhar, Felix Laumann, and Marcus Liwicki.
\newblock A comprehensive guide to {B}ayesian convolutional neural network with
  variational inference.
\newblock {\em arXiv preprint arXiv:1901.02731}, 2019.

\bibitem{blundell2015}
Charles Blundell, Julien Cornebise, Koray Kavukcuoglu, and Daan Wierstra.
\newblock Weight uncertainty in neural network.
\newblock In {\em International conference on machine learning}, pages
  1613--1622. PMLR, 2015.

\bibitem{kruegeret2018}
D.~Krueger, C.W. Huang, R.~Islam, R.~Turner, and A.~Courville A.~Lacoste.
\newblock {B}ayesian hypernetworks.
\newblock {\em https://arxiv.org/abs/1710.04759v2}, 2018.

\bibitem{tran2019}
D.~Tran, M.~Dusenberry, and D.~Hafner M.~van~der Wilk.
\newblock {B}ayesian layers: A module for neural network uncertainty.
\newblock {\em Conference on Neural Information Processing Systems (NeurIPS
  2019)}, 1313:14660–14672, 2019.

\bibitem{szegedy2013}
Christian Szegedy, Wojciech Zaremba, Ilya Sutskever, Joan Bruna, Dumitru Erhan,
  Ian Goodfellow, and Rob Fergus.
\newblock Intriguing properties of neural networks.
\newblock {\em arXiv preprint arXiv:1312.6199}, 2013.

\bibitem{vadera2022methods}
Sunil Vadera and Salem Ameen.
\newblock Methods for pruning deep neural networks.
\newblock {\em IEEE Access}, 10:63280--63300, 2022.

\bibitem{hassibi1993}
B.~Hassibi, D.G. Stork, and G.J. Wolff.
\newblock Optimal brain surgeon and general network pruning.
\newblock In {\em IEEE International Conference on Neural Networks}, pages
  293--299 vol.1, 1993.

\bibitem{srivastava2014dropout}
Nitish Srivastava, Geoffrey~E. Hinton, Alex Krizhevsky, Ilya Sutskever, and
  Ruslan Salakhutdinov.
\newblock Dropout: a simple way to prevent neural networks from overfitting.
\newblock 15(1):1929--1958, 2014.

\bibitem{wan2013regularization}
Li~Wan, Matthew Zeiler, Sixin Zhang, Yann Le~Cun, and Rob Fergus.
\newblock Regularization of neural networks using dropconnect.
\newblock In {\em International conference on machine learning}, pages
  1058--1066. PMLR, 2013.

\bibitem{sakai2019}
Yasufumi Sakai, Bruno~U. Pedroni, Siddharth Joshi, Abraham Akinin, and Gert
  Cauwenberghs.
\newblock Dropout and dropconnect for reliable neuromorphic inference under
  energy and bandwidth constraints in network connectivity.
\newblock In {\em 2019 IEEE International Conference on Artificial Intelligence
  Circuits and Systems (AICAS)}, pages 76--80, 2019.

\bibitem{chan2020unlabelled}
Alex Chan, Ahmed Alaa, Zhaozhi Qian, and Mihaela Van Der~Schaar.
\newblock Unlabelled data improves {B}ayesian uncertainty calibration under
  covariate shift.
\newblock In {\em International conference on machine learning}, pages
  1392--1402. PMLR, 2020.

\bibitem{zhou2022sparse}
Hongpeng Zhou, Chahine Ibrahim, Wei~Xing Zheng, and Wei Pan.
\newblock Sparse {B}ayesian deep learning for dynamic system identification.
\newblock {\em Automatica}, 144:110489, 2022.

\bibitem{sandhu2020model}
Rimple Sandhu.
\newblock {\em Model comparison and sparse learning of nonlinear physics-based
  models using {B}ayesian inference}.
\newblock PhD thesis, Carleton University, 2020.

\bibitem{sandhu2021}
Rimple Sandhu, Mohammad Khalil, Chris Pettit, Dominique Poirel, and Abhijit
  Sarkar.
\newblock Nonlinear sparse {B}ayesian learning for physics-based models.
\newblock {\em Journal of Computational Physics}, 426:109728, 2021.

\bibitem{dabiran_jcp2023}
Nastaran Dabiran, Brandon Robinson, Rimple Sandhu, Mohammad Khalil, Chris~L.
  Pettit, Dominique Poirel, and Abhijit Sarkar.
\newblock Exploring hierarchical framework of nonlinear sparse {B}ayesian
  learning algorithm through numerical investigations.
\newblock {\em https://arxiv.org/pdf/2310.14749.pdf}, 2023.

\bibitem{schwobel2022last}
Pola Schw{\"o}bel, Martin J{\o}rgensen, Sebastian~W Ober, and Mark Van
  Der~Wilk.
\newblock Last layer marginal likelihood for invariance learning.
\newblock In {\em International Conference on Artificial Intelligence and
  Statistics}, pages 3542--3555. PMLR, 2022.

\bibitem{wilson2016deep}
Andrew~Gordon Wilson, Zhiting Hu, Ruslan Salakhutdinov, and Eric~P Xing.
\newblock Deep kernel learning.
\newblock In {\em Artificial intelligence and statistics}, pages 370--378.
  PMLR, 2016.

\bibitem{van2018learning}
Mark van~der Wilk, Matthias Bauer, ST~John, and James Hensman.
\newblock Learning invariances using the marginal likelihood.
\newblock {\em Advances in Neural Information Processing Systems}, 31, 2018.

\bibitem{hinton2006fast}
Geoffrey~E Hinton, Simon Osindero, and Yee-Whye Teh.
\newblock A fast learning algorithm for deep belief nets.
\newblock {\em Neural computation}, 18(7):1527--1554, 2006.

\bibitem{bengio2006greedy}
Yoshua Bengio, Pascal Lamblin, Dan Popovici, and Hugo Larochelle.
\newblock Greedy layer-wise training of deep networks.
\newblock {\em Advances in neural information processing systems}, 19, 2006.

\bibitem{bridgman2023robust}
Wyatt Bridgman, Reese~E Jones, and Mohammad Khalil.
\newblock Robust scalable initialization for {B}ayesian variational inference
  with multi-modal {L}aplace approximations.
\newblock {\em Probabilistic Engineering Mechanics}, page 103540, 2023.

\bibitem{higham2019deep}
Catherine~F Higham and Desmond~J Higham.
\newblock Deep learning: An introduction for applied mathematicians.
\newblock {\em SIAM review}, 61(4):860--891, 2019.

\bibitem{higham2019}
Catherine~F Higham and Desmond~J Higham.
\newblock Deep learning: An introduction for applied mathematicians.
\newblock {\em SIAM review}, 61(4):860--891, 2019.

\bibitem{fink1997compendium}
Daniel Fink.
\newblock A compendium of conjugate priors.
\newblock Technical report, 1997.

\bibitem{guo2017calibration}
Chuan Guo, Geoff Pleiss, Yu~Sun, and Kilian~Q Weinberger.
\newblock On calibration of modern neural networks.
\newblock In {\em International conference on machine learning}, pages
  1321--1330. PMLR, 2017.

\bibitem{nixon2019measuring}
Jeremy Nixon, Michael~W Dusenberry, Linchuan Zhang, Ghassen Jerfel, and Dustin
  Tran.
\newblock Measuring calibration in deep learning.
\newblock In {\em CVPR workshops}, volume~2, 2019.

\bibitem{tipping2001}
Michael~E Tipping.
\newblock Sparse {B}ayesian learning and the relevance vector machine.
\newblock {\em Journal of machine learning research}, 1(Jun):211--244, 2001.

\bibitem{gilks1995markov}
W.R. Gilks, S.~Richardson, and D.~Spiegelhalter.
\newblock {\em Markov Chain {M}onte {C}arlo in {P}ractice}.
\newblock Chapman \& Hall/CRC Interdisciplinary Statistics. Taylor \& Francis,
  1995.

\bibitem{ching2007}
Y.~Chen J.~Ching.
\newblock Transitional {M}arkov {C}hain {M}onte {C}arlo method for {B}ayesian
  model updating, model class selection, and model averaging.
\newblock {\em Journal of Engineering Mechanics}, 133 (7)(816–832), 2007.

\bibitem{drugowitsch2013}
Jan Drugowitsch.
\newblock Variational {B}ayesian inference for linear and logistic regression.
\newblock {\em arXiv preprint arXiv:1310.5438}, 2013.

\bibitem{jeffreys1946}
Harold Jeffreys.
\newblock An invariant form for the prior probability in estimation problems.
\newblock {\em Proceedings of the Royal Society of London. Series A,
  Mathematical and Physical Sciences}, 186(1007):453--461, 1946.

\bibitem{tipping1999}
Michael Tipping.
\newblock The relevance vector machine.
\newblock {\em Advances in neural information processing systems}, 12, 1999.

\bibitem{sandhu2017}
Rimple Sandhu, Chris Pettit, Mohammad Khalil, Dominique Poirel, and Abhijit
  Sarkar.
\newblock {B}ayesian model selection using automatic relevance determination
  for nonlinear dynamical systems.
\newblock {\em Computer Methods in Applied Mechanics and Engineering},
  320:237--260, 2017.

\bibitem{sandhu2022}
Rimple Sandhu, Brandon Robinson, Mohammad Khalil, Chris~L Pettit, Dominique
  Poirel, and Abhijit Sarkar.
\newblock Encoding nonlinear and unsteady aerodynamics of limit cycle
  oscillations using nonlinear sparse {B}ayesian learning.
\newblock {\em arXiv preprint arXiv:2210.11476}, 2022.

\bibitem{conn2000trust}
Andrew~R Conn, Nicholas~IM Gould, and Philippe~L Toint.
\newblock {\em Trust region methods}.
\newblock SIAM, 2000.

\bibitem{bishop2006}
Christopher~M Bishop and Nasser~M Nasrabadi.
\newblock {\em Pattern recognition and machine learning}, volume~4.
\newblock Springer, 2006.

\bibitem{butler2018}
T.~Butler, J.~Jakeman, and T.~Wildey.
\newblock Combining push-forward measures and {B}ayes' rule to construct
  consistent solutions to stochastic inverse problems.
\newblock {\em SIAM Journal on Scientific Computing}, 40(2):A984--A1011, 2018.

\end{thebibliography}

\end{document}